\documentclass[a4paper,11pt]{article}
\pdfoutput=1 

\usepackage{jinstpub} 
\usepackage{subcaption}
\usepackage{caption}
\usepackage{xcolor}
\usepackage[subrefformat=parens,labelformat=parens]{subcaption}

\title{Numerical Qualification of Eco-Friendly Gas Mixtures for Avalanche-Mode Operation of Resistive Plate Chambers in INO-ICAL}

\author[a,b]{Jaydeep Datta,\note{Corresponding author.}}
\author[a,b]{Sridhar Tripathy}
\author[a,b]{Nayana Majumdar}
\author[a,b]{and Supratik Mukhopadhyay}

\affiliation[a]{Saha Institute of Nuclear Physics, Sector 1, AF Block, Bidhan Nagar, Salt Lake, Kolkata 700064, India}
\affiliation[b]{Homi Bhabha National Institute, Training School Complex, Anushaktinagar, Mumbai 400094, India}

\emailAdd{jaydeep.datta@gmail.com}

\abstract{Proposition and qualification of an eco-friendly gas mixture of Ar(5\%):CO$_2$(60\%): N$_2$(35\%) for operating Resistive Plate Chambers of INO-ICAL experiment in avalanche mode is the soul of this work. The detector efficiency and streamer probability for the proposed mixture have been simulated using a hydrodynamic model developed by us. Efficacy of the mixture has been studied by comparing the simulated results to available experimental data for the standard gas mixture of R134a(95.2\%):i-C$_4$H$_{10}$(4.5\%):SF$_6$(0.3\%) to be used in INO-ICAL as well as the same observables measured for other eco-friendly hydrofluoroolefin (HFO1234ze)-based potential replacements. To reduce the streamer probability of the proposed argon-based gas mixture at higher voltages, the usual practice of adding a small amount of electronegative gas SF$_6$ has been tested. An alternative approach of lowering electronic threshold has been investigated as well for the same purpose.}


\arxivnumber{2103.01906} 

\begin{document}
\maketitle
\flushbottom
\section{\label{sec:1}Introduction}
Resistive Plate Chamber (RPC) is a gaseous detector that has found wide applications in fundamental physics experiments. Excellent time and fairly good position resolution together with ease of construction of large detection area at comparatively lower cost have facilitated its utilization in many large-scale experiments. The magnetized Iron Calorimeter (ICAL) at India-based Neutrino Observatory (INO) \cite{INO2017} is one such setup that is contemplated as a detector for studying atmospheric neutrinos. The major prospects of the ICAL detector will be improvement in the precision of neutrino oscillation parameters by observing one full oscillation period and studying the matter-dependent effects on neutrino properties for assessing the sign of mass-squared difference, $\Delta_{32}$, and magnitude of mixing angle, $\theta_{23}$, both of which are as yet unknown. The ICAL is designed as a tracking calorimeter containing a large target mass (50 kt of iron) to achieve statistically significant number of neutrino interactions in a reasonable time frame. It will be equipped with RPCs ($\sim$ 29,000) interspersed with horizontal layers of iron plates for tracking the muons produced in charged-current interaction of neutrinos with the iron nuclei. The charge identification of the muons (to distinguish between neutrino and its anti-particle) will be accomplished by application of a high magnetic field ($\sim$ 1.5 T) across the ICAL. The position and timing information of the muons will be recorded by the RPCs during their flight through the calorimeter which will lead to reconstruction of their flight path and directionality from the time-of-flight measurement. These data would be utilized for determining energy, $E$, and path-length, $L$, of the neutrino and anti-neutrino separately, which are important for detecting the oscillation pattern in the $L/E$ spectrum. 

\textcolor{red}{The ICAL tracking calorimeter will use single-gap RPCs made of commercially available glass of thickness 3 mm with gas gap of 2 mm and an active area of dimension 2 m $\times$ 2 m.
The resistivity of the glass material to be used as the electrodes is $\sim10^{12}~\Omega-\rm m$.
The RPCs will be operated in avalanche mode with a gas mixture of 1,1,1,2-tetrafluoroethane, C$_2$H$_2$F$_4$(95.2\%) (commercially known as R134a) and isobutane, i-C$_4$H$_{10}$ (4.5\%) with a very small amount of sulfur hexafluoride, SF$_6$ (0.3\%).
At any instance the ICAL setup populated with about 29,000 RPCs will require a gas volume of 216 m$^3$.}
However, the Global Warming Potential (GWP) of the R134a and the SF$_6$ are 1300 and 23900, respectively, along with that of i-C$_4$H$_{10}$ being 3. It makes the effective GWP of the said gas mixture little more than 1300 which is well beyond the permissible limit of 150, set by the Kyoto protocol \cite{Kyoto}, adopted in 1997 by United Nations Framework Convention on Climate Change (UNFCCC) in order to limit and reduce greenhouse gas emission.
\textcolor{red}{Although a closed loop gas circulation system will be used in ICAL, the possibility of small leaks can not be ruled out which may turn out substantial for such a large volume of gas. For the present choice of gas mixture having a high GWP of about 1300, this can be an issue of concern.} The issue certainly necessitates exploration of alternative gas mixtures with sufficiently low GWP for operating the RPCs without compromising the objectives of the experiments. 

The problem has already been addressed with extensive studies on various gases and their properties in order to find suitable substitutes for operating RPCs in many other high-energy physics experiments \cite{Abbrescia2016, Guida2016, Saviano2018, Bianchi2019, Bianchi2020, Guida:2020lrs, Abbrescia:2016xdh}. According to these studies, one allotropic form of tetrafluoropropene, C$_3$H$_2$F$_4$ (commercially known as HFO1234ze) which is close to R134a in chemical structure with very low GWP ($\sim$ 6), can be regarded as a potential replacement. However, the HFO1234ze shows lower effective Townsend coefficient with respect to that of R134a when compared at same electric field. As a result, comparatively higher electric field is required to achieve an efficiency of about 90\%. A few promising mixtures based on HFO1234ze, i-C$_4$H$_{10}$ and He or CO$_2$ could be identified with low GWP (< 150) \cite{Abbrescia2016, Bianchi2020} for efficient operation of RPC though the high-voltage requirement of some of them \cite{Bianchi2020} might not be compatible to ICAL high-voltage supply system \cite{Manna:2018lem}. Moreover, these mixtures have shown presence of substantially large fraction of streamers \cite{Abbrescia2016, Bianchi2020} at higher efficiency which is well above the tolerable limit for safe and long-term operation. As a result, inclusion of small amount of electron quencher gas SF$_6$ ($\sim$ 1$\%$) in the mixtures could not be ruled out for effective operation of the detectors. This approach has led the effective GWP sometimes cross the limit of 150 \cite{Abbrescia2016, Bianchi2019, Guida:2020lrs, Abbrescia:2016xdh}. Other than this, HFO1234ze has one more disadvantage regarding its use in RPC considering the criticism about its degrading effect on the detector health. It has low GWP essentially because of its smaller lifetime in the environment. The rate of dissociation of the gas in presence of OH$^-$ radical (produced from dissociation of water molecule) is four times faster than that of R134a \cite{NorwayReport}. The final products are HF and other corrosive acids which can destroy the surface of the RPC electrodes. This turns out to be detrimental for neutrino experiments planned to operate for a long time in which good health of the detector and its consistent performance are of crucial importance.

The previous studies along with a need for eco-friendly RPC operation in INO-ICAL have motivated us to investigate for an alternate feasible solution. In this context, we have developed a numerical model on the basis of hydrodynamics following the idea of P. Fonte \cite{Fonte2013} for simulating the performance of RPC operated with any gas mixture. The model computes transportation and the Townsend multiplication of primary charges produced due to ionization of gaseous molecules following the transport equations of hydrodynamics for a specific electric field. Depending upon evolution of the charges, one can identify the working mode of the RPC as avalanche or streamer and the corresponding operating voltage regimes for a given gas mixture. This model is expected to provide us a handle to qualify any proposed gas mixture for its application in RPC operation as per requirements of the experiment. 

The model has been implemented on the platform of COMSOL Multiphysics \cite{Comsol}, a commercial package of Finite Element Method (FEM). A few other packages, namely, HEED \cite{Heed2005} and MAGBOLTZ \cite{Magboltz1999}, have been utilized to obtain the supplementary information on primary ionization and electron transport parameters for the gas mixture under study, respectively. The model has been developed and validated by comparing its results for a gas mixture of R134a and n-C$_4$H$_{10}$ with reported experimental data of the detector efficiency and streamer probability at different applied voltages \cite{Camarri1998}. This work has been reported by us in \cite{Jaydeep2020}.

In the present work, we have compared the simulated values of the efficiency and streamer probability obtained for the standard R134a-based gas mixture with the experimental data reported in \cite{Abbrescia:2016xdh, Guida:2020lrs} to validate the numerical model. The same model has been used to qualify a mixture of argon (Ar)(5\%), carbon dioxide (CO$_2$)(60\%) and nitrogen (N$_2$)(35\%) proposed as a non-inflammable, eco-friendly as well as cost-effective solution for operating RPCs of the INO-ICAL setup in avalanche mode. The simulated efficiency and streamer probability as a function of high-voltage supply for the said mixture have been compared to the same observables measured with other HFO1234ze-based potential mixtures proposed in \cite{Abbrescia2016, Bianchi2020}.

\section{\label{sec:2}Numerical Model}
In gaseous ionization detectors, the filling gas usually comprises of two individual gas components. The principal component acts as the medium of Townsend ionization and hence, constitutes larger portion of the mixture. The passage of charged particles through the detector gas volume may lead to the generation of primary pairs of electron and ion through ionization and excitation of gas molecules. The de-excitation process releases photons which induce additional ionization, namely photo-ionization, along with the Townsend one. So, a second poly-atomic component is included in the mixture to serve as a quencher of the photons and thus reduce the contribution of photo-ionization.

The present hydrodynamic model has considered the gas mixture as a charged solution. It assumes the neutral gas molecules as solvent and all the charge species (electrons and ions) produced in the medium due to ionization as well as photo-ionization as solutes. Since the concentration of the charges is much less than that of the neutral gas molecules, the transport of the charges has been modeled using the \emph{"Transport of Dilute Species"} module of COMSOL. The following drift-diffusion-reaction equation \ref{transeqn1} has been solved to compute the field-dependent charge transportation. 
\begin{eqnarray}
\label{transeqn1}
\frac{\partial n_k}{\partial t} + \vec{\nabla} \cdot (-D_k \vec{\nabla} n_k + \vec {u}_k n_k) = R_k\\
\label{transeqn2}
R_k = S_e + S_{ph}\\
\label{transeqn3}
S_e = (\alpha (\vec{E}) - \eta (\vec{E})) |\vec{u}_e| n_e (\vec{x}, t)\\ 
 \label{transeqn4}
S_{ph} = Q_e \mu_{abs} \psi_0
\end{eqnarray}
where $n_k$, $k= i, e$, represents the concentration of the ions and electrons, respectively, while $D_k$, $\vec u_k$ and $R_k$ are their diffusion, drift velocity and rate of production, respectively. As is obvious from equation \ref{transeqn2}, $R_k$, is the sum of two source terms, $S_e$ and $S_{ph}$, of the charges produced through Townsend ionization and photo-ionization mechanisms, respectively. The source term due to Townsend ionization will be same for both the electrons and ions. According to equation \ref{transeqn3}, $S_e$ is dependent upon the transport parameters of electrons, namely, the first Townsend coefficient, $\alpha$, attachment coefficient, $\eta$, and drift velocity, $\vec {u_e}$. The parameters as a function of the electric field have been calculated using MAGBOLTZ for the given gas mixture under study. The $S_{ph}$ can be computed using equation \ref{transeqn4} where $Q_e$ is quantum efficiency of the filling gas mixture for electron generation from photo-ionization, $\mu_{abs}$ is photo-absorption coefficient of the quencher, and $\psi_0$ is the photon flux generated in the detection volume. The $\mu_{abs}$ has been calculated considering the corresponding photo-absorption cross-section of the specific gas component obtained from relevant sources \cite{Lombos1967, Orlando1991}. To determine $\psi_0$, a diffusion-like approximation of the photon propagation in the gas medium, as shown in equation \ref{phprop1}, has been considered following Capeill{\`{e}}re et al. \cite{Capeillere2008}. It has been computed using \emph{"Coefficient Form Partial Differential Equation"} module of COMSOL.
\begin{eqnarray}
\label{phprop1}          
\vec{\nabla} (-c\vec{\nabla} \psi_0) + a \psi_0 = f\\  
\label{phprop2}
c = \frac{1}{3 \mu_{abs}}\\
\label{phprop3}
a = \mu_{abs}\\
\label{phprop4}
f = \delta S_e
\end{eqnarray}
Here, $\delta$ in equation \ref{phprop4} represents the number of excited molecules for each ionized molecule. 

It is well known that the slow movement of ions gives rise to subsequent development of space charge in the detector that can distort the applied electric field. So, the electric field, $\vec E$, has been calculated at small time steps taking into account the space charge density, $\rho$, using the following equations \ref{efield1} and \ref{efield2}. The \emph{"Electrostatic"} module of COMSOL has been used for this purpose.
\begin{eqnarray}
\label{efield1}
\vec{E}=-\vec{\nabla} V\\
\label{efield2}
-\vec{\nabla} (\epsilon_0 \vec{\nabla} V - \vec{P}) = \rho\\
\rho = q_e(n_i -n_e)
\end{eqnarray}
Here, $V$ is the potential difference, $\vec{P}$ is the polarization vector, $q_e$ is the magnitude of the electronic charge and $\epsilon_0$ is the permittivity of the vacuum.

The following boundary conditions have been used in the model. The loss of electrons from the gas gap upon reaching the anode has been taken care of by assuming drift of the electrons through the anode. Similar condition has been set about the ions to consider their outflow at the cathode. To incorporate the phenomenon of electron and ion diffusing and drifting out of the simulated volume, the two boundaries other than the cathode and the anode have been assumed open for them.
The photon flux at the electrodes has been taken as zero as these electrodes are made up of materials which do not have scintillating property. This condition has been implemented as a Dirichlet boundary condition in the model. The photon propagation out of the simulation volume has been taken into account by considering the two boundaries other than the electrodes open for them.

\section{\label{sec:Gas}Choice of Gas Mixture}
In the present work, a gas mixture of Ar(5\%):CO$_2$(60\%):N$_2$(35\%) has been proposed as a new alternative for avalanche mode operation of RPCs in INO-ICAL. It should be noted that all the components other than CO$_2$ (GWP = 1) are green (GWP = 0). Moreover, all the components are and commercially available and economical as well.

The argon gas has been taken as the principal component for ionization. Its high Townsend coefficient at low electric field will help to reduce the operating voltage of RPCs, while on the other hand, will lead to large streamer probability at higher fields. To circumvent the issue of streamer, the partial volume of the Ar has been reduced by introducing N$_2$, which is electronegative and chemically less reactive. As a photon quencher, we have preferred CO$_2$ to other alternatives as it has less penning transfer coefficient in a mixture with Ar \cite{Sahin:2010ssz} and less GWP \cite{Saviano2018}. Its non-flammability and cheap cost are added advantages.

\begin{figure}[h]	 
 \centering
 \subcaptionbox{\label{fig:1a}}[0.49\linewidth]{\includegraphics[width=\linewidth]{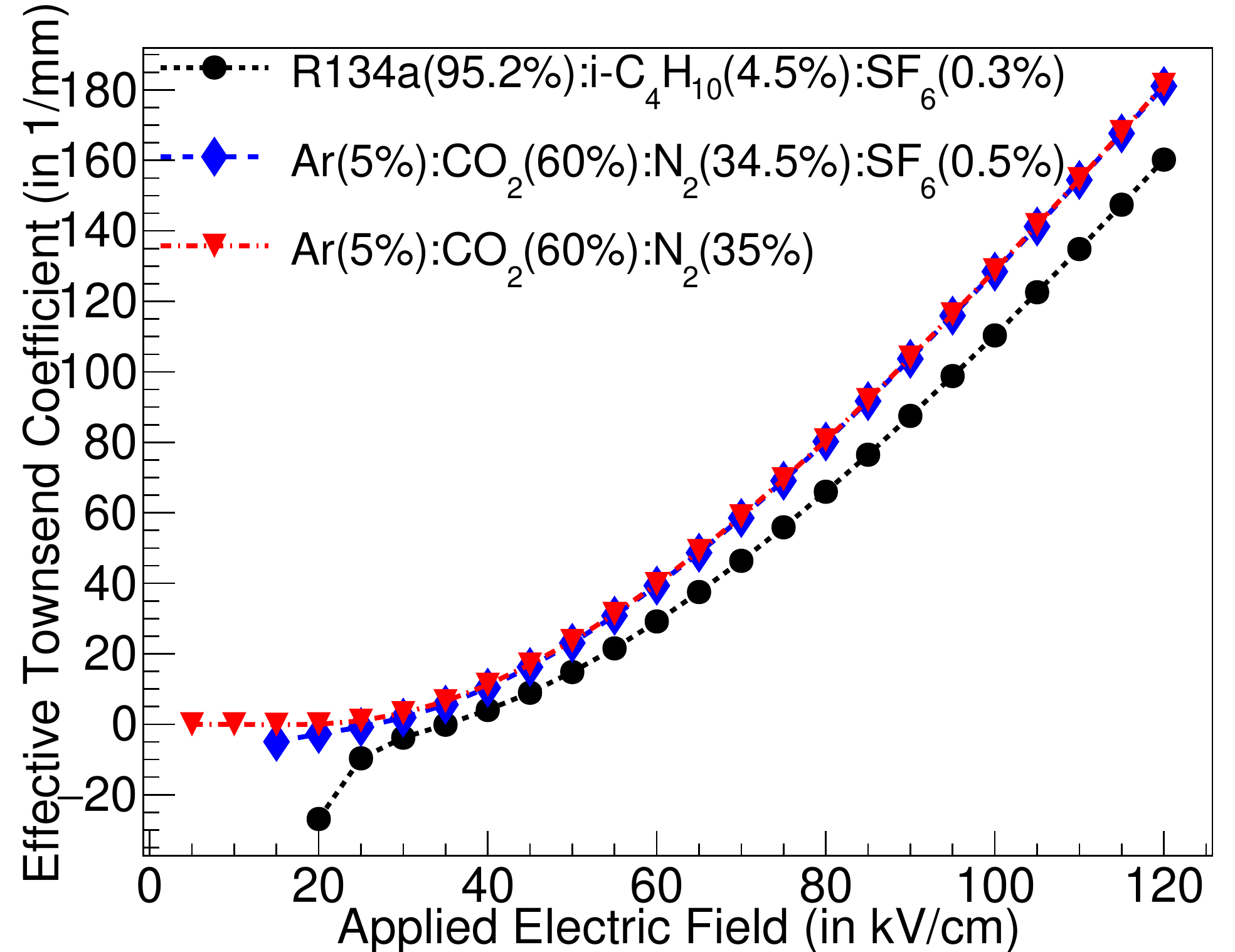}} 
 \subcaptionbox{\label{fig:1b}}[0.49\linewidth]{\includegraphics[width=\linewidth]{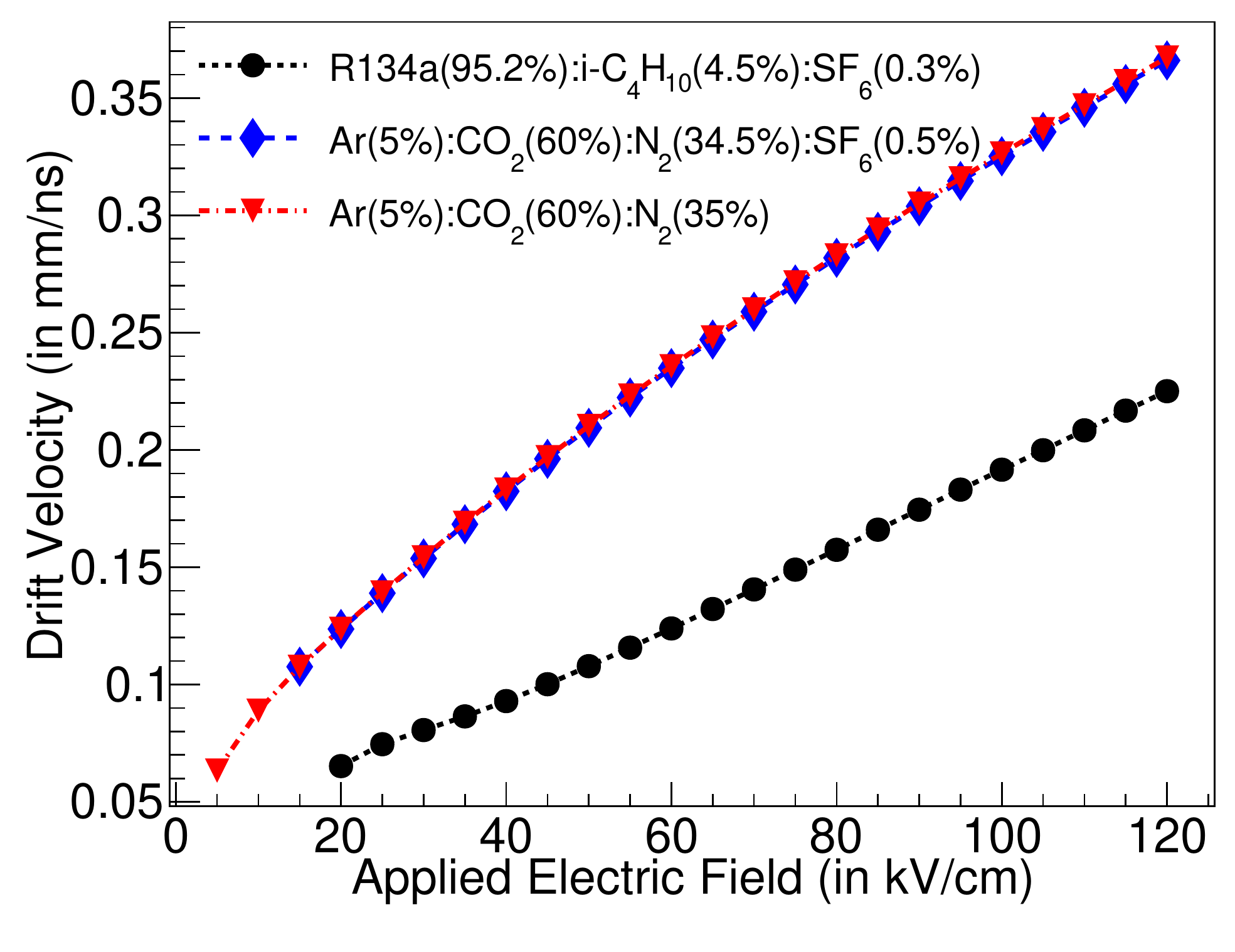}}
 \caption{Comparison of (a) effective Townsend coefficient and (b) drift velocity as a function of applied voltage of proposed Ar(5\%):CO$_2$(60\%):N$_2$(35\%) and Ar(5$\%$):CO$_2$(60\%):N$_2$ (34.5\%):SF$_{6}$(0.5\%) mixture to that of R134a(95.2\%):i-C$_4$H$_{10}$(4.5\%):SF$_6$(0.3\%) as obtained from MAGBOLTZ \cite{Magboltz1999} calculated for temperature 293.15 K and 1 bar of pressure.}
 \label{fig:1}
\end{figure}

\begin{figure}[h]	 
	\centering
	\includegraphics[width=0.6\linewidth]{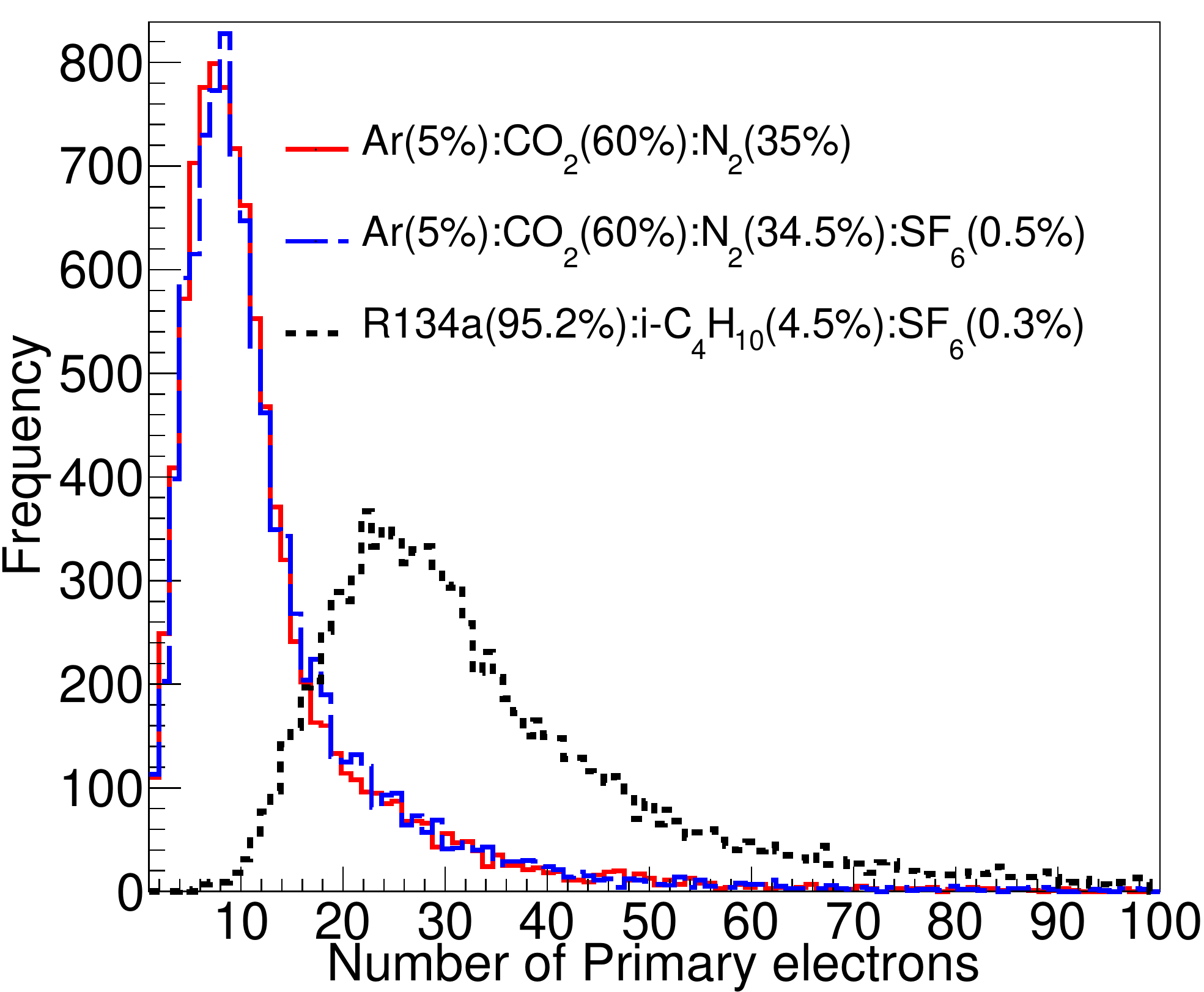}
	\caption{Comparison of primary electron number (for 10000 events) produced in proposed Ar(5\%):CO$_2$(60\%):N$_2$(35\%) and Ar(5$\%$):CO$_2$(60\%):N$_2$(34.5\%):SF$_{6}$(0.5\%) mixture to that of R134a(95.2\%):i-C$_4$H$_{10}$(4.5\%):SF$_6$(0.3\%) as obtained from HEED \cite{Heed2005}. The values have been calculated assuming the gas temperature is 293.15K and pressure 1 bar.}
	\label{fig:2}
\end{figure}

To study the performance of the proposed Ar-based gas mixture, a comparison of its transport parameters to that of the standard R134a-based gas mixture has been made with the help of MAGBOLTZ. The effect of adding a small amount of SF$_{6}$(0.5\%), a widely used electron quencher, to the new mixture has been studied as well. The results of the effective Townsend coefficient and drift velocity as a function of the applied electric field for all these gas mixtures have been plotted in figure \ref{fig:1}. It should be noted that the pressure and temperature have been kept at 1 bar and 293.15 K, respectively, throughout the numerical work. It is obvious from the plot shown in figure \ref{fig:1a} that the proposed Ar-based gas mixture has closely followed the standard R134a-based one in the case of effective Townsend coefficient which indicates similar Townsend multiplication of electrons in both the mixtures. 
The comparison of drift velocity, as depicted in figure \ref{fig:1b}, shows that the proposed Ar-based mixture is faster by a factor of 2 than the standard R134a-based one. However, the mean primary cluster size is smaller in it by a factor of 3 in comparison to that of the standard R134a-based mixture, as calculated from HEED and shown in figure \ref{fig:2}.
According to Ramo's theorem \cite{Ramo}, the induced current at an instant $t$ on the RPC readout can be determined following the equation \ref{Ramo1}.
\begin{eqnarray}\label{Ramo1}
	i(t)=-\vec{W} \cdot \vec{u}(t) q_e n_e(t)\\
	|\vec{W}| =\frac{\epsilon_r}{2b+d\epsilon_r}\label{Ramo2}
\end{eqnarray}
where $n_e$ and $q_e$ are the electron number and the charge, respectively, $\vec u$ is the drift velocity and $\vec W$ is the weighting field. \textcolor{red}{It has been calculated following equation \ref{Ramo2} which is obtained for a parallel-plate geometry assuming the pickup electrodes much larger than the gas gap and electrode thickness \cite{Riegler:2002vg}. In equation \ref{Ramo2}, b and d denotes the thickness of the resistive electrode and gas gap, respectively, and the $\epsilon_r$ is the relative permittivity of the resistive material.} The higher drift velocity of the new mixture has compensated the lack of primary electrons which determines the total number of electrons, $n_e$, for a given Townsend multiplication. It implies that the current signal produced with the proposed Ar-based mixture should be comparable with that obtained using the standard R134a-based mixture. It should also be noted that the addition of SF$_6$ by a small amount (0.5\%) to the proposed gas does not alter the transport properties significantly. While calculating the electron transport properties using MAGBOLTZ, we have kept the temperature and pressure inputs same as that of the standard R134a-based gas mixture. We have also included penning transfer coefficient following \cite{Sahin:2017wyp, Sahin:2014haa} as this mixture can have penning effect. 

\section{\label{sec:simproc}Simulation Procedure}
The present work involves the simulation of dynamics and evolution of the charge species produced due to passage of cosmic muons using the hydrodynamic model in a 2D geometry of RPC filled with a gas mixture. It has been performed for different applied voltages and the corresponding efficiency and streamer probability have been estimated. Different aspects of the calculation procedure have been discussed in the following sub-sections.

\subsection{\label{sec:geo}Geometry}
\begin{figure}
	\centering
	\includegraphics[width=0.7\linewidth]{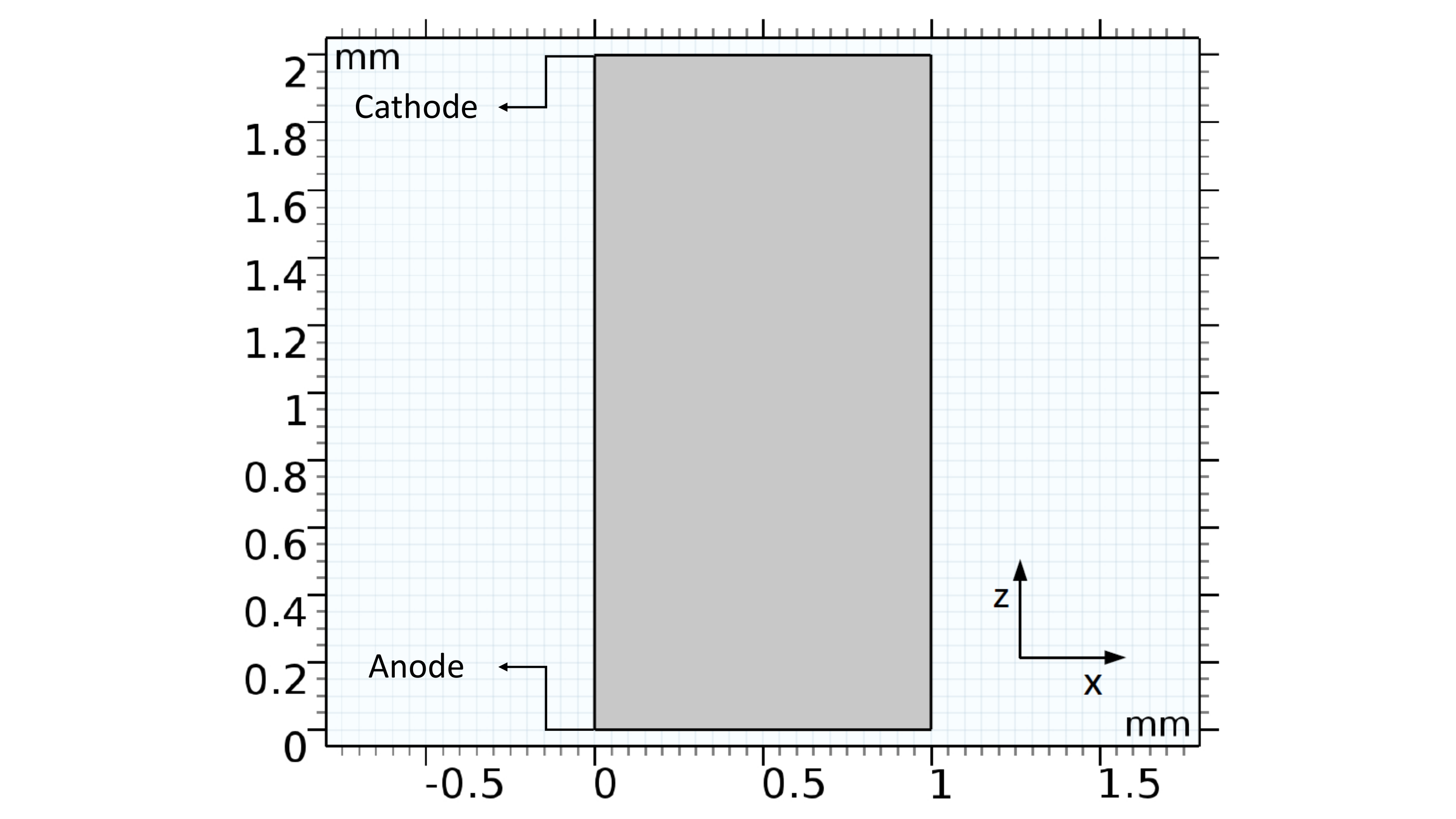}
	\caption{2D geometry of RPC used in this work for simulation}
	\label{fig:geom}
\end{figure}

The model of the RPC has been built in 2D Cartesian geometry (XZ-plane) with a gas gap of 2 mm along the Z-direction following the design parameters of the RPCs of the INO-ICAL as shown in figure \ref{fig:geom}.
Throughout the simulation we have considered the applied electric field is along the positive Z-direction. So, we have anode at Z = 0 mm and cathode at Z = 2 mm. In order to limit the computation, the length along the X-direction has been kept 1 mm which is sufficient to contain the spread of the avalanche \cite{Jaydeep2020}. It is conceivable that the 2D geometry has its own limitations and 3D simulation should give us a more realistic situation. However, we have opted for the 2D geometry to optimize between heavy computational expense and reliability of the results. It has been shown in \cite{Ammosov:1996gg} that assuming the absence of any spacer, the voltage across the gas gap of an RPC is the same as that applied on the resistive electrodes. Throughout the work, we have considered that the applied voltage across the electrodes is seen by the gas gap without any drop.

\subsection{\label{sec:evgen}Event Generation}
In the simulation, 10,000 cosmic muons with energy between 1 - 10 GeV with zenith angle 0$^o$ - 13$^{o}$ have been considered. We have followed cosmic muon flux given by Tang et al \cite{Tang} and used HEED to compute the number of primary electrons generated in the gas gap due to these muon events. For each event, it has provided the number of primary clusters, their position and number of electrons in each cluster. It has been found that for a muon event, the clusters are distributed throughout the whole gas gap and in most cases contain one electron. In the simulation, the primary clusters produced in each muon event have been collectively represented as a single cluster with its position and size (number of electrons in it) calculated from the information provided by HEED. The Z-position of the same has been obtained from the weighted mean of the position values of the primary clusters and the size from the sum of all the primary electrons.

\begin{figure}[h]	 
	\centering
	\subcaptionbox{\label{fig:3a}}[0.49\linewidth]{\includegraphics[width=\linewidth]{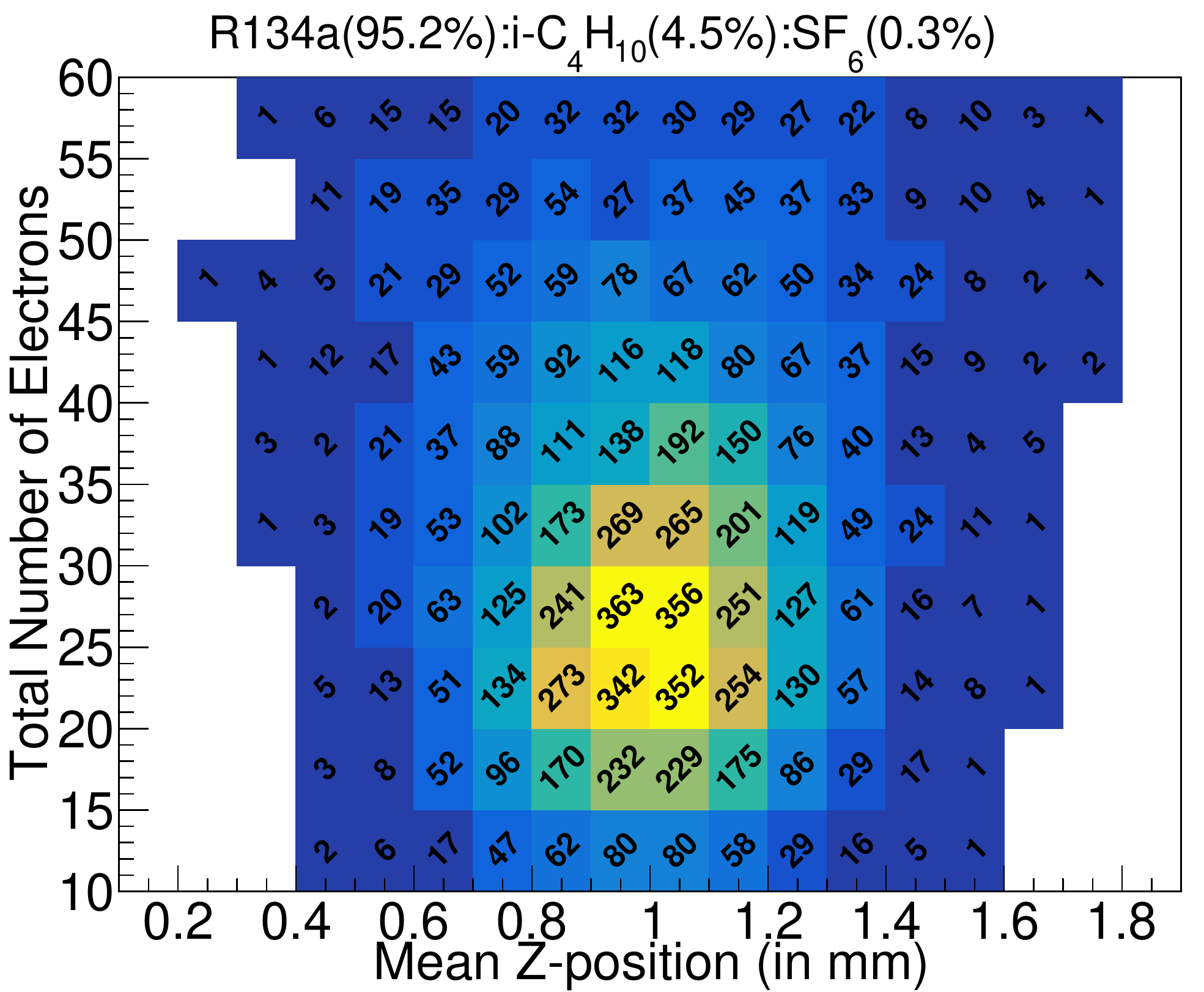}} 
 \subcaptionbox{\label{fig:3b}}[0.49\linewidth]{\includegraphics[width=\linewidth]{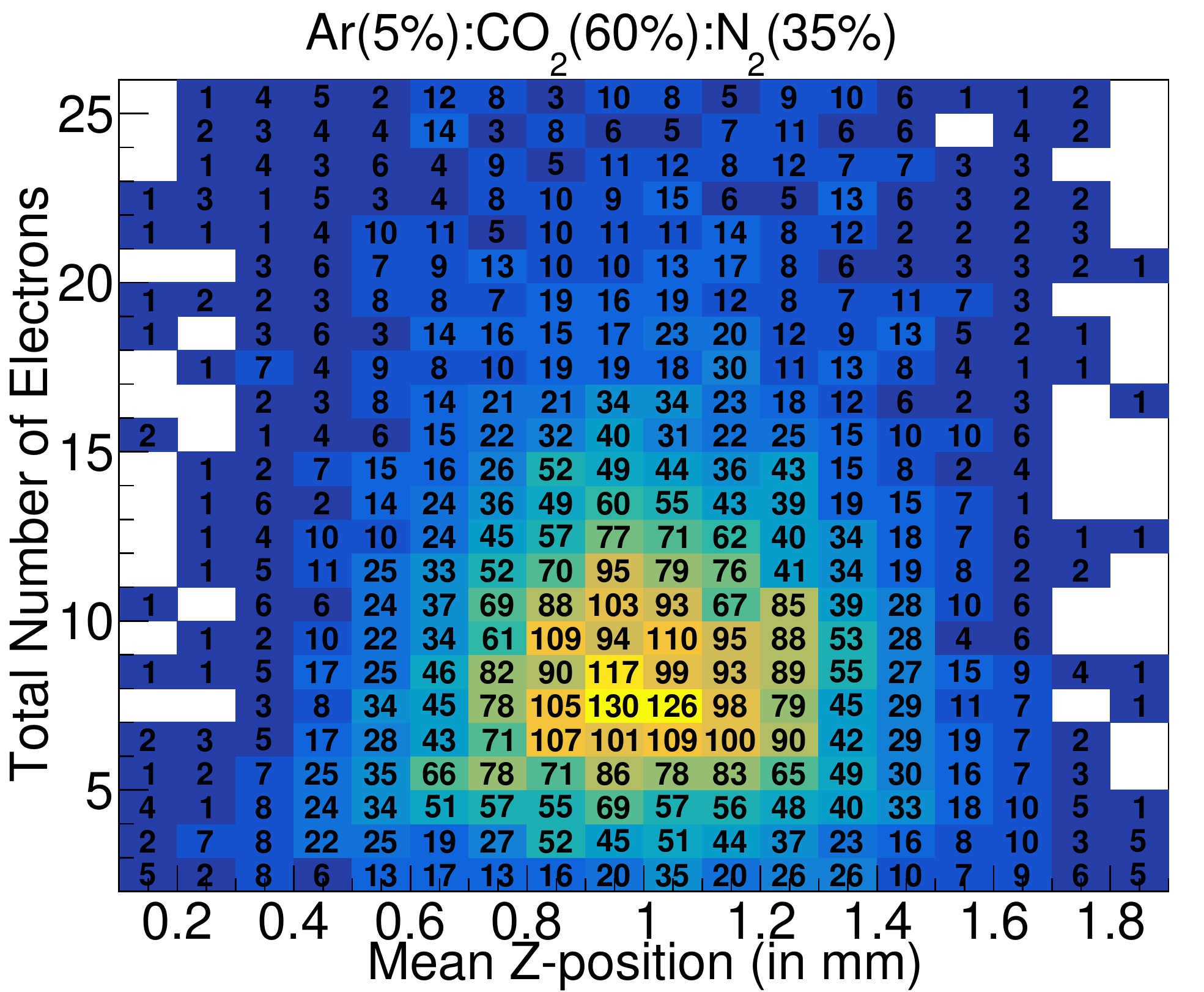}}
	\caption{2D histogram of muon events for the gas mixture of (a) R134a(95.2\%):i-C$_4$H$_{10}$ (4.5\%):SF$_6$(0.3\%) and (b) Ar(5\%):CO$_2$(60\%):N$_2$(35\%)}
	\label{fig:3}
\end{figure}

In figure \ref{fig:3a}, a 2D histogram of primary electrons for the standard R134a-based gas mixture has been shown assuming the bin size for the mean Z-position of the clusters to be 0.1 mm while that for the cluster size as 5. The number in each cell denotes the number of events which have clusters having mean Z-position and their size falling within the range of that cell. For the gas mixture R134a(95.2\%):i-C$_4$H$_{10}$(4.5\%):SF$_6$(0.3\%), the average primary electron is higher than 20 and 90\% of them lie between 10 and 60. In principle, one should look at all the events and consider their contribution. But due to the fact that the events which have fewer electrons than 10 and created very near to the anode will not produce either a streamer or an avalanche large enough to be detected by the electronics available, we can leave them out of the calculation without affecting the result much so as to save the computational expense. The other events which have higher number of total electrons can be generated at any position in the gas gap. It has been observed from the HEED data that a significant number of events will produce clusters very close to the anode making them unsuitable for streamer formation and in some cases for valid event also. We have carried out the calculation for events which have cluster size larger than 60 and have been generated away from the anode. It increases the computational time without affecting the result. Therefore, for R134a-based gas mixture, if we consider 90\% of the total events (events with total number of primaries in the range 10 to 60), we will be able to achieve a reasonable solution with less computational expense and marginal error.

Similar 2D histogram of the primary clusters for the gas mixture Ar(5\%):CO$_2$(60\%):N$_2$(35\%) has been shown in figure \ref{fig:3b} following the HEED data. In this case, we have observed that for 90\% events, the cluster size has varied between 2 - 25. So, the 2D histogram has been built with a smaller bin size of 1 for the cluster size while keeping the bin size for mean Z-position unaltered from the previous case.

\subsection{\label{sec:detres}Estimation of Detector Responses}
It is apparent from the discussion in section \ref{sec:evgen} that for each of the 10,000 muon events, a seed cluster having definite size and mean Z-position could be prepared to initiate the hydrodynamic simulation of detector response for a given voltage configuration. From the numerical study, a correlation of the occurrence of avalanche or streamer at an applied voltage to the size and the position of the seed cluster has been observed. It implies that the avalanche and streamer events are viable to be distinguished on the basis of these two parameters. Once the least cluster size to develop into a streamer for a given mean position is determined, it is obvious that all the events with larger clusters at the same position should lead to streamer. Similar condition could be identified for the mean position for a given cluster size. Using these observations, we could identify the events from the 2D histogram leading to either of the avalanche or streamer without carrying out the hydrodynamic simulation for all the events. It has facilitated the estimation of the total number of events leading to either of the avalanche or streamer summing up the events identified as avalanche or streamer from the 2D histograms. Subsequently, the streamer probability has been calculated by dividing the number of streamer events with the total number of muon events. This method of calculation has helped us to reduce the computational expense and time.

Calculation of the detector efficiency has been done following the same way. Only those events that have produced signals crossing a predefined threshold of current, have been considered as valid events. In experiments, signal in terms of voltage is collected after electronic amplification of the current signal across a load resistance. In our simulation, we have calculated the current equivalent to the voltage mentioned as threshold after the electronic gain. The ratio of the valid events with respect to the total number of events has been regarded as the efficiency. The detailed discussion on the method of calculation of efficiency and streamer probability may be found in our previous work \cite{Jaydeep2020}. 

\subsection{\label{sec:stp}Stopping Conditions}
The classification of events leading to either of streamer and avalanche has been done on the basis of following conditions.
\begin{itemize}
	\item {\bf Avalanche:} All the electrons have been collected in the anode.
	\item {\bf Streamer:} When the space charge field has become equal to the applied field ($\gamma = 1$)
\end{itemize}
where $\gamma = (E_{\rm total}-E_{\rm applied})/{E_{\rm applied}}$ is a parameter for comparing the space charge field to the applied one. Here, E$_{\rm applied}$ represents the applied electric field while E$_{\rm total}$ is the total electric field considering the field produced by the space charge along with the actual one.. 

\section{\label{sec:result}Results}
The discussion of the results of the present work has been divided into two parts. In section \ref{sec:standard}, we have shown the numerical estimates of RPC efficiency and streamer probability for the standard mixture of R134a(95.2\%):i-C$_4$H$_{10}$(4.5\%):SF$_6$(0.3\%) to be used in the RPCs of INO-ICAL. The results have been compared to reported experimental measurements \cite{Guida:2020lrs, Abbrescia:2016xdh} to demonstrate the validity of the model. 

In section \ref{sec:propose}, the same observables have been calculated for the proposed Ar(5\%):CO$_2$(60\%): N$_2(35\%)$ gas mixture to study its efficacy for its application in the RPCs of INO-ICAL. These results have been compared to the performance of other potential eco-friendly mixtures \cite{Abbrescia2016, Bianchi2020} as well to envisage the figures of merit of the present gas mixture.

\subsection{\label{sec:standard}Standard Gas Mixture} 
For this gas mixture, we have considered 3 mm thick electrode with relative permittivity 6.25 while calculating the weighting field \textcolor{red}{following equation \ref{Ramo1}. The signal calculated using Ramo’s theorem provides an intrinsic response of the detector without any external electronic contribution. However, from the experiment with glass RPC, a threshold of 5 mV has been found adequate to discriminate the signal from noise after an amplification of 80. To implement the said threshold, the current equivalence of the same has been calculated for the induced signal has been obtained in terms of current. As the voltage threshold of 5 mV is actually 0.0625 mV without the gain factor, it is equivalent to a current of 1.25 $\mu$A across 50 $\Omega$ termination. Since the current signal has been calculated for a model geometry with 1 mm length in X-direction, the threshold can be considered to be applicable for 1 mm pickup strip as the length in Y-direction can be considered of any length in a 2D geometry.}

\begin{figure}[h]	 
	\centering
	\subcaptionbox{\label{fig:4a}}[0.49\linewidth]{\includegraphics[width=\linewidth]{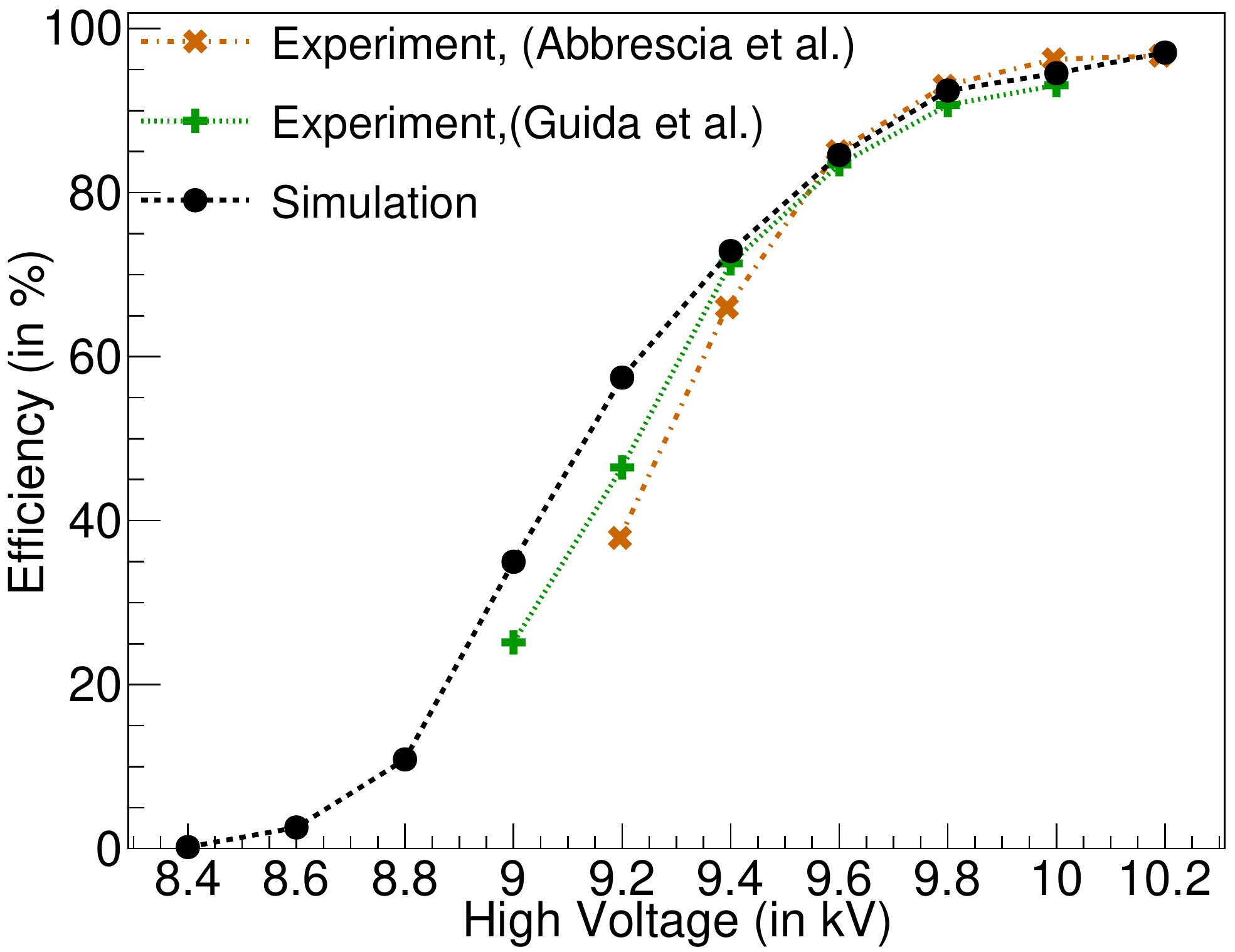}} 
 \subcaptionbox{\label{fig:4b}}[0.49\linewidth]{\includegraphics[width=\linewidth]{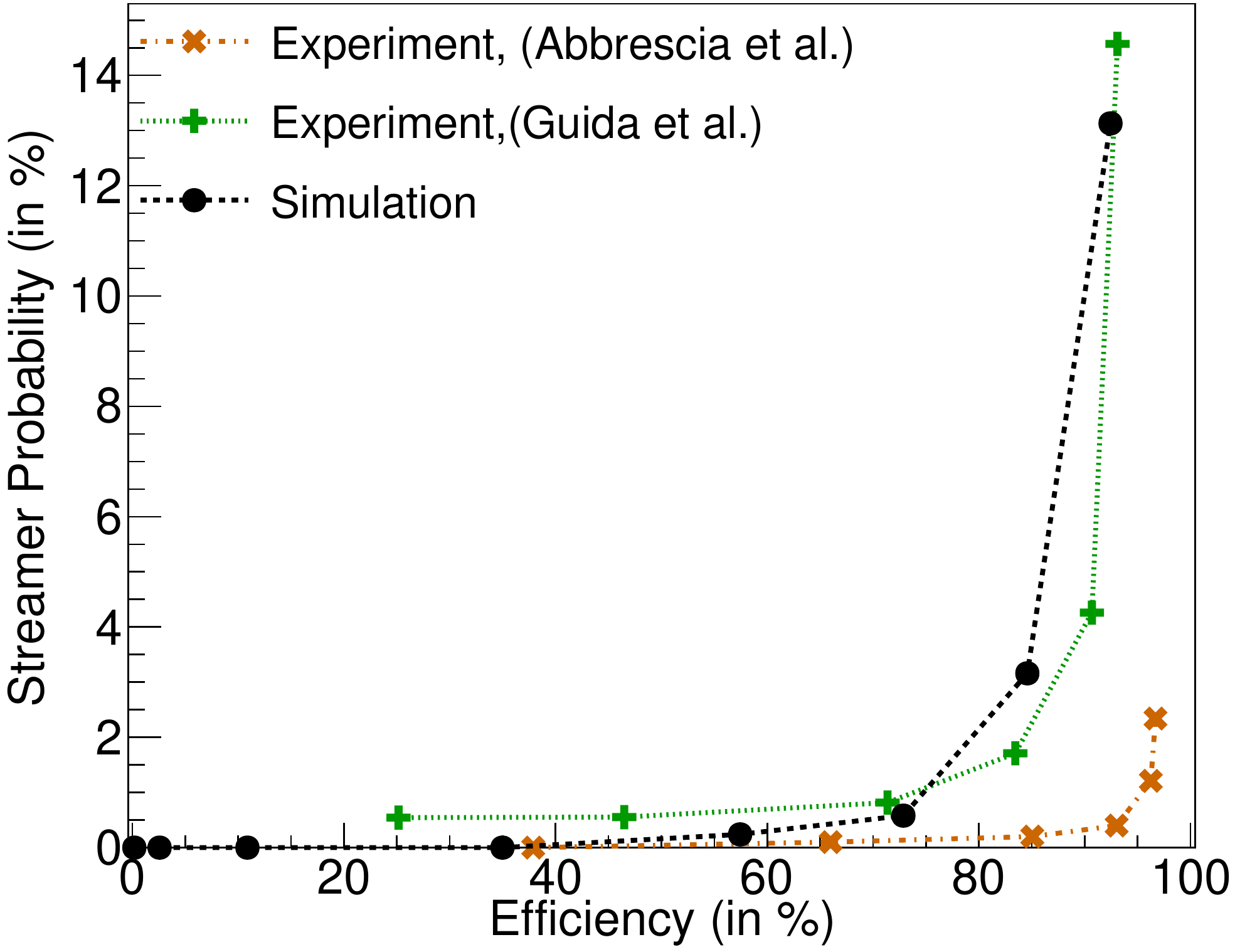}}
	\caption{Comparison between simulation (this work) and experiments \cite{Abbrescia:2016xdh, Guida:2020lrs} (a) efficiency as function of high-voltage, (b) streamer probability as a function of efficiency of R134a(95.2\%):i-C$_4$H$_{10}$(4.5\%):SF$_6$(0.3\%)}
	\label{fig:4}
\end{figure}

The calculated efficiency as a function of high-voltage and the calculated streamer probability as a function of efficiency have been shown in figures \ref{fig:4a} and \ref{fig:4b}, respectively, for the standard mixture of R134a(95.2\%):i-C$_4$H$_{10}$(4.5\%):SF$_6$(0.3\%). Comparison with a few measurements \cite{Guida:2020lrs,Abbrescia:2016xdh} in the same plots have shown a close agreement between the simulation and the experiments.

\subsection{\label{sec:propose}Proposed Gas Mixtures} 
\begin{figure}[h]	 
	\centering
	\includegraphics[width=0.5\linewidth]{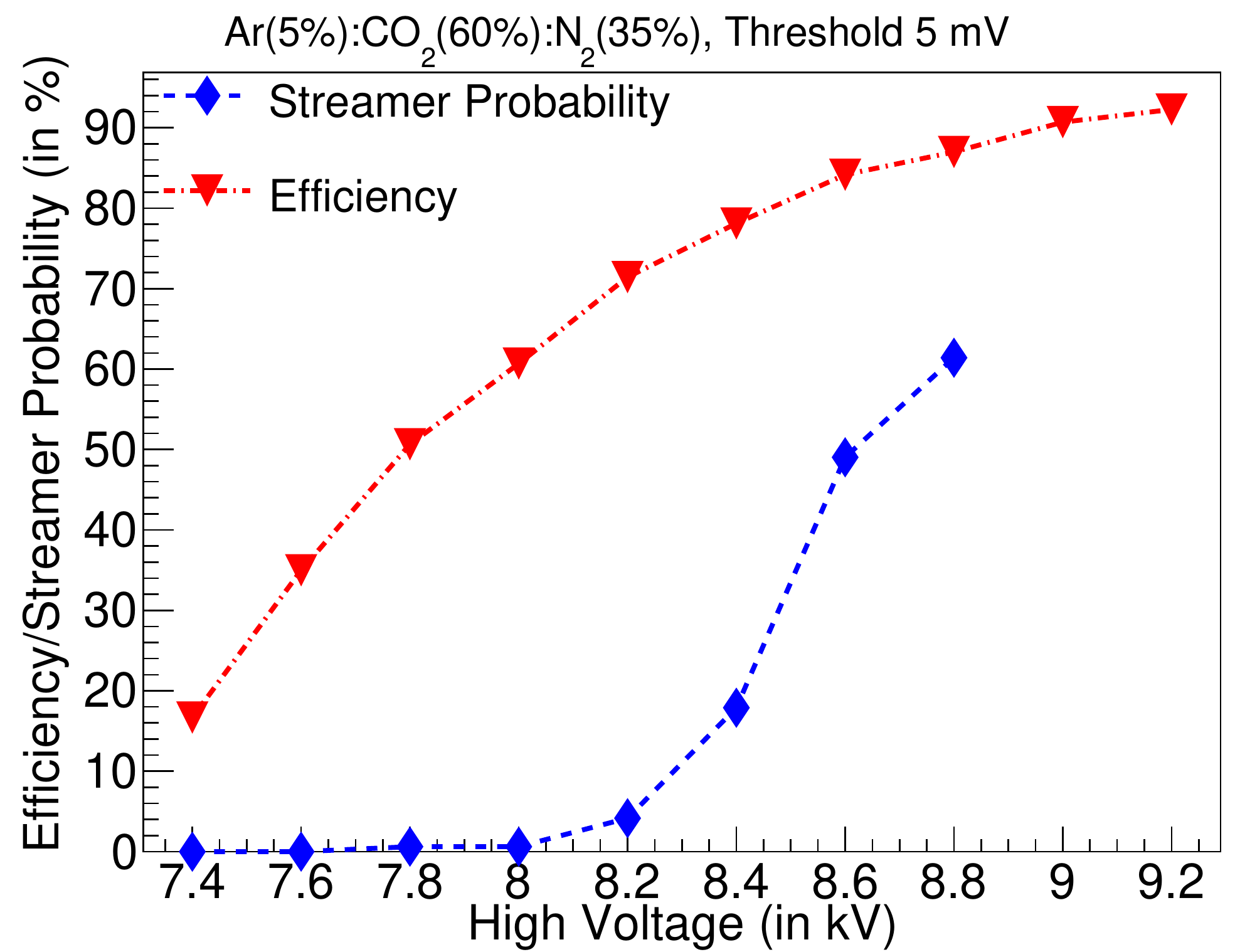}
 	\caption{Simulated streamer probability and efficiency as functions of the applied high-voltage for Ar(5\%):CO$_2$(60\%):N$_2$(35\%) with threshold of 5 mV.}
	\label{fig:5}
\end{figure}

\begin{figure}[h]	 
	\centering
	\subcaptionbox{\label{fig:6a}}[0.49\linewidth]{\includegraphics[width=\linewidth]{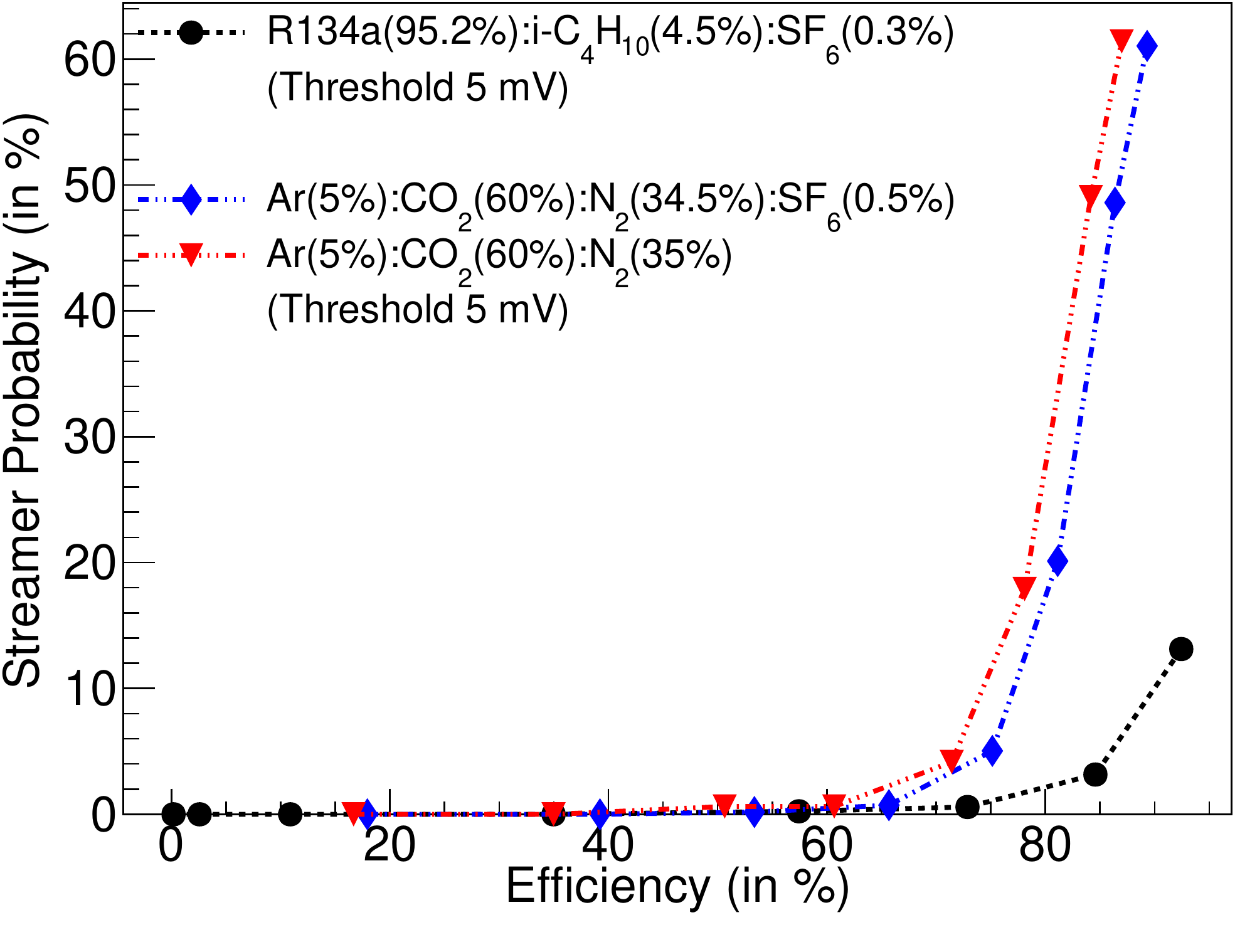}}
	\subcaptionbox{\label{fig:6b}}[0.49\linewidth]{\includegraphics[width=\linewidth]{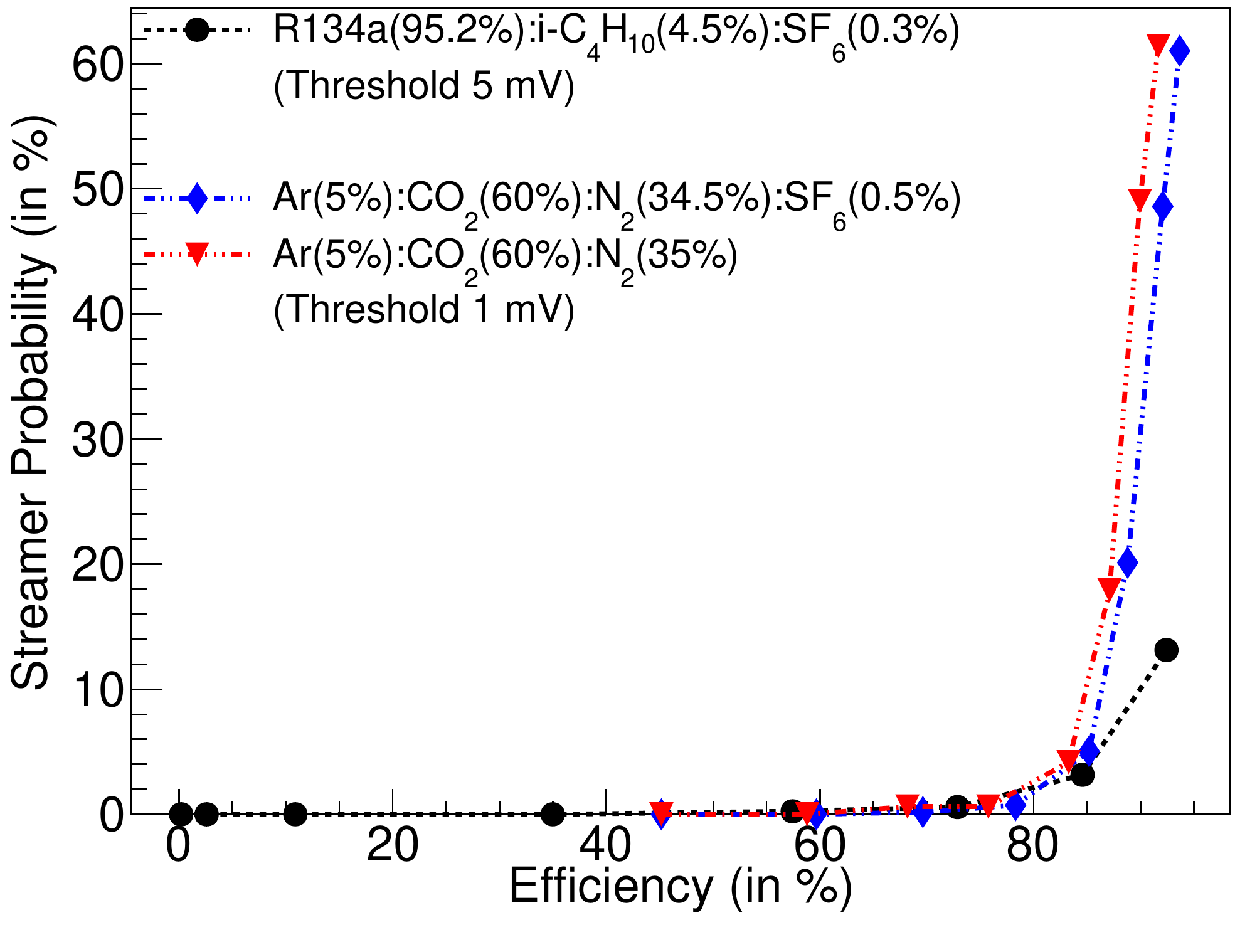}}
	\caption{Comparison of streamer probability as function of efficiency among R134a(95.2\%):i-C$_4$H$_{10}$(4.5\%):SF$_6$(0.3\%), Ar(5\%):CO$_2$(60\%):N$_2$(35\%) and Ar(5\%):CO$_2$(60\%):N$_2$(34.5\%):SF$_6$ (0.5\%) mixtures for the threshold of (a) 5 mV and (b) 1mV.}
	\label{fig:6}
\end{figure}

Efficiency and streamer probability of the proposed Ar-based mixture with variation of applied voltage have been depicted in figure \ref{fig:5}. It has been observed that the streamer probability of the said mixture is fairly large at higher efficiency. For example, it has shot close to 50\% at 85\% efficiency which is not suitable for efficient operation of RPC in avalanche mode.
We have noted that this problem can be alleviated in two ways.
The first option is to reduce the streamer probability by addition of very small amount of electronegative SF$_6$ as practiced in case of the standard R134a-based mixture.
The GWP of the Ar-based mixture will be within the limit of 150 if SF$_6$ is added by a percentage of less than 0.5\%.
\textcolor{red}{The other option is to reduce the threshold for valid events which improves the ratio of avalanche to streamer events leading to an increase in efficiency although it can allow some noise eventually.}
Both the options have been studied and the expected improvement has been observed in both the cases. The effect of adding 0.5\% of SF$_6$ in the mixture has been depicted in figure \ref{fig:6a} where the efficiency of the mixture with and without the SF$_6$ component has been compared to that of the standard mixture. It can be noted from the figure that the operation in avalanche mode with the Ar-based mixture is not practically feasible beyond 80-85\% efficiency. The addition of SF$_6$ component in the mixture has reduced the streamer probability to around 20\% when efficiency is close to 90\%, although it is still higher in comparison to the standard R134a-based mixture. The results of reducing the threshold from 5 mV to 1 mV have been depicted in figure \ref{fig:6b} where an overall reduction in the streamer probability of the Ar-based mixture is visible. This approach can lead to an acceptable streamer probability (< 10\%) around 85\% efficiency which is comparable to that of the standard R134a-based mixture. It can be noted that by reducing the threshold, the present Ar-based mixture can operate efficiently with low streamer probability even without the SF$_6$ component. Moreover, it should be noted from figure \ref{fig:5} that the proposed mixture has allowed the RPC to be operated at much lower voltages which is obviously an important advantage of this mixture. 

\begin{figure}[h]	 
	\centering
	\subcaptionbox{\label{fig:7a}}[0.49\linewidth]{\includegraphics[width=\linewidth]{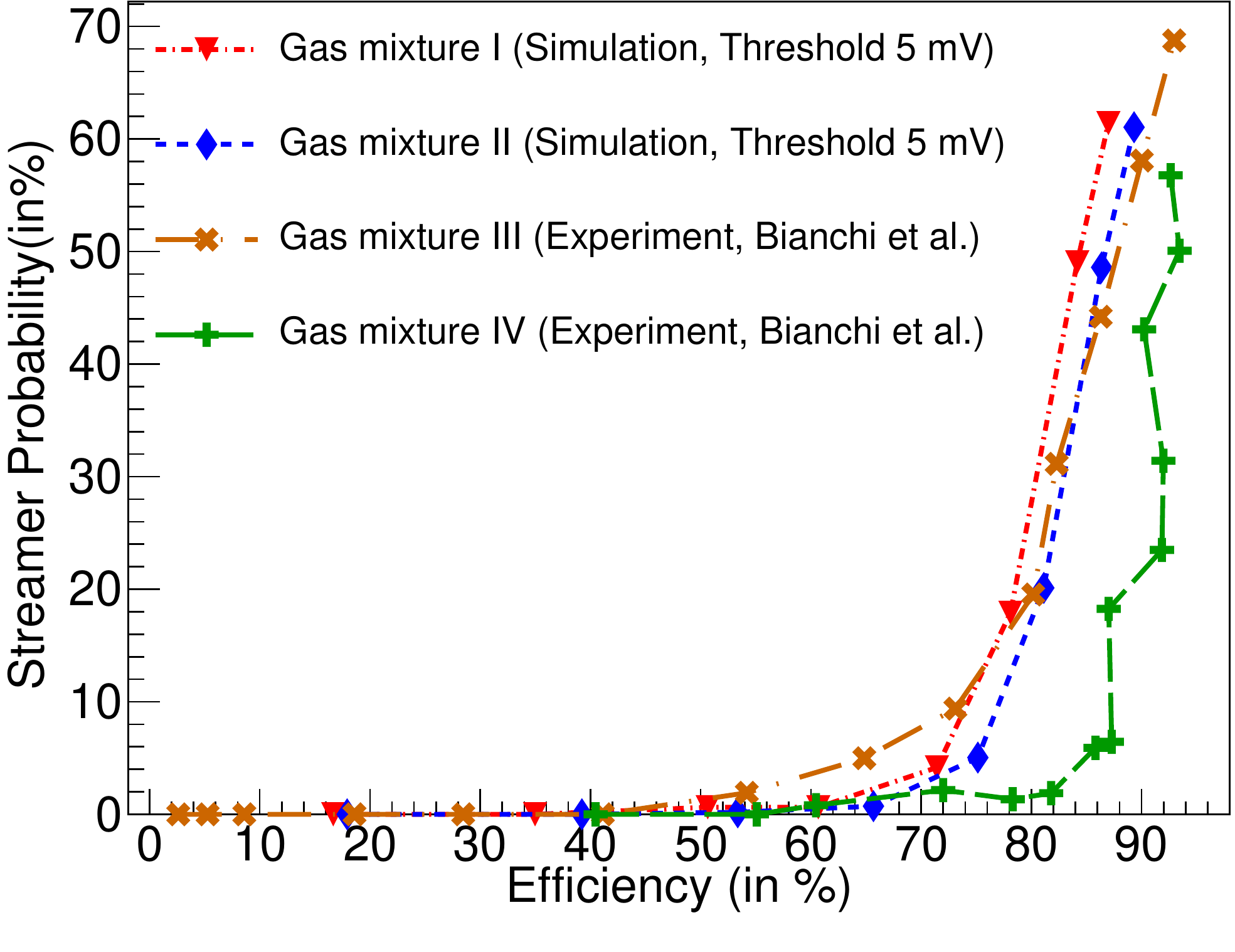}}
	\subcaptionbox{\label{fig:7b}}[0.49\linewidth]{\includegraphics[width=\linewidth]{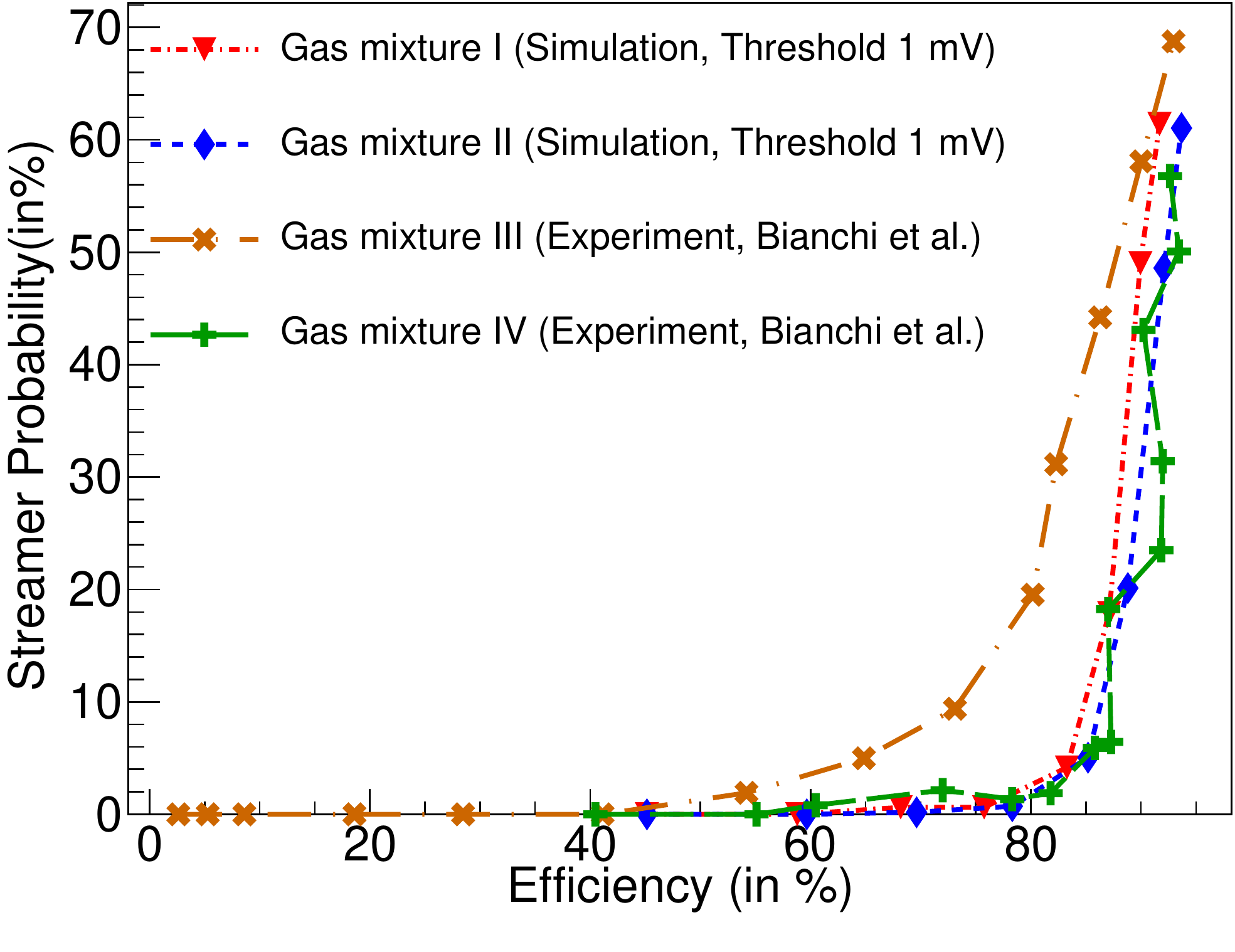}}\\
	\subcaptionbox{\label{fig:7c}}[0.49\linewidth]{\includegraphics[width=\linewidth]{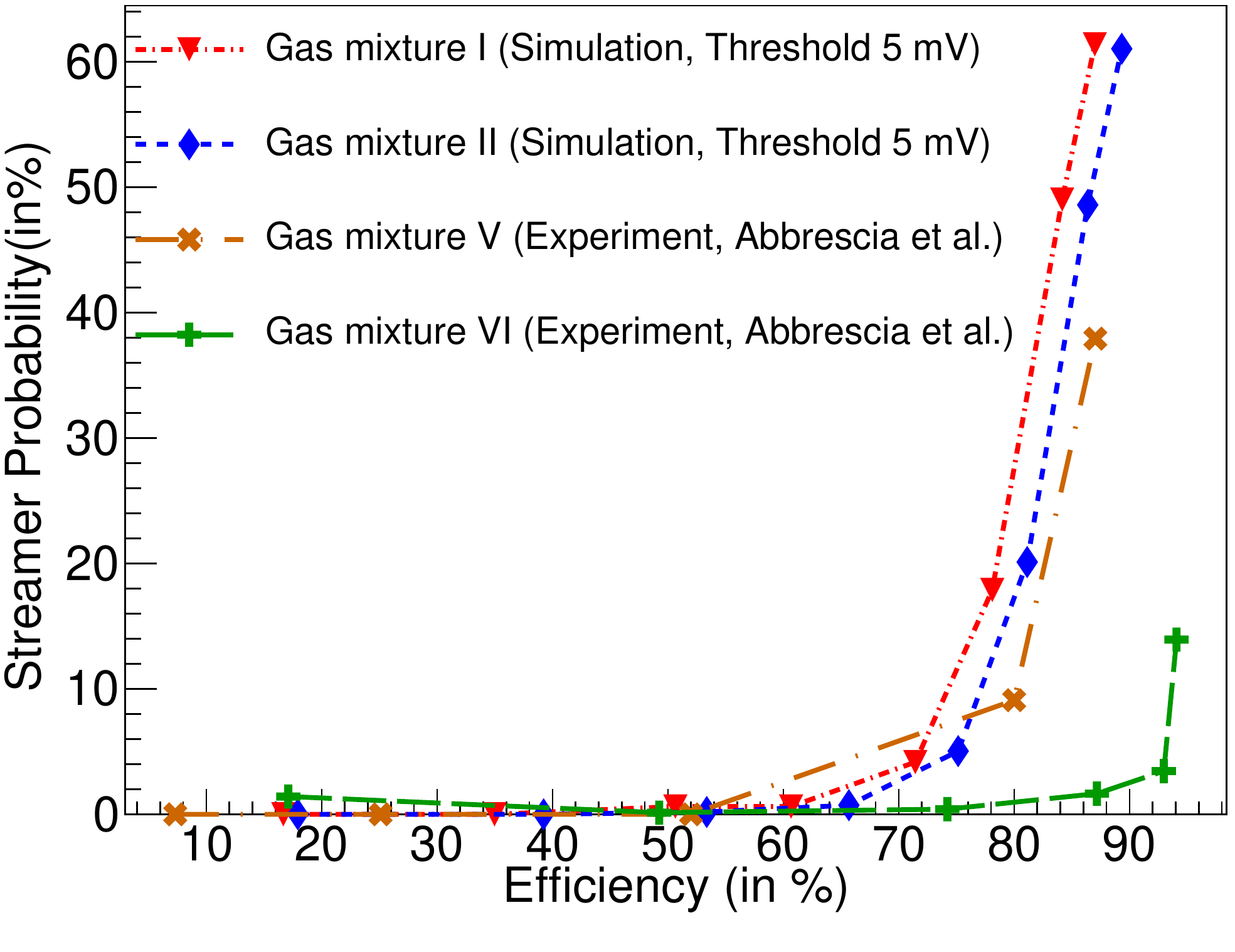}}
	\subcaptionbox{\label{fig:7d}}[0.49\linewidth]{\includegraphics[width=\linewidth]{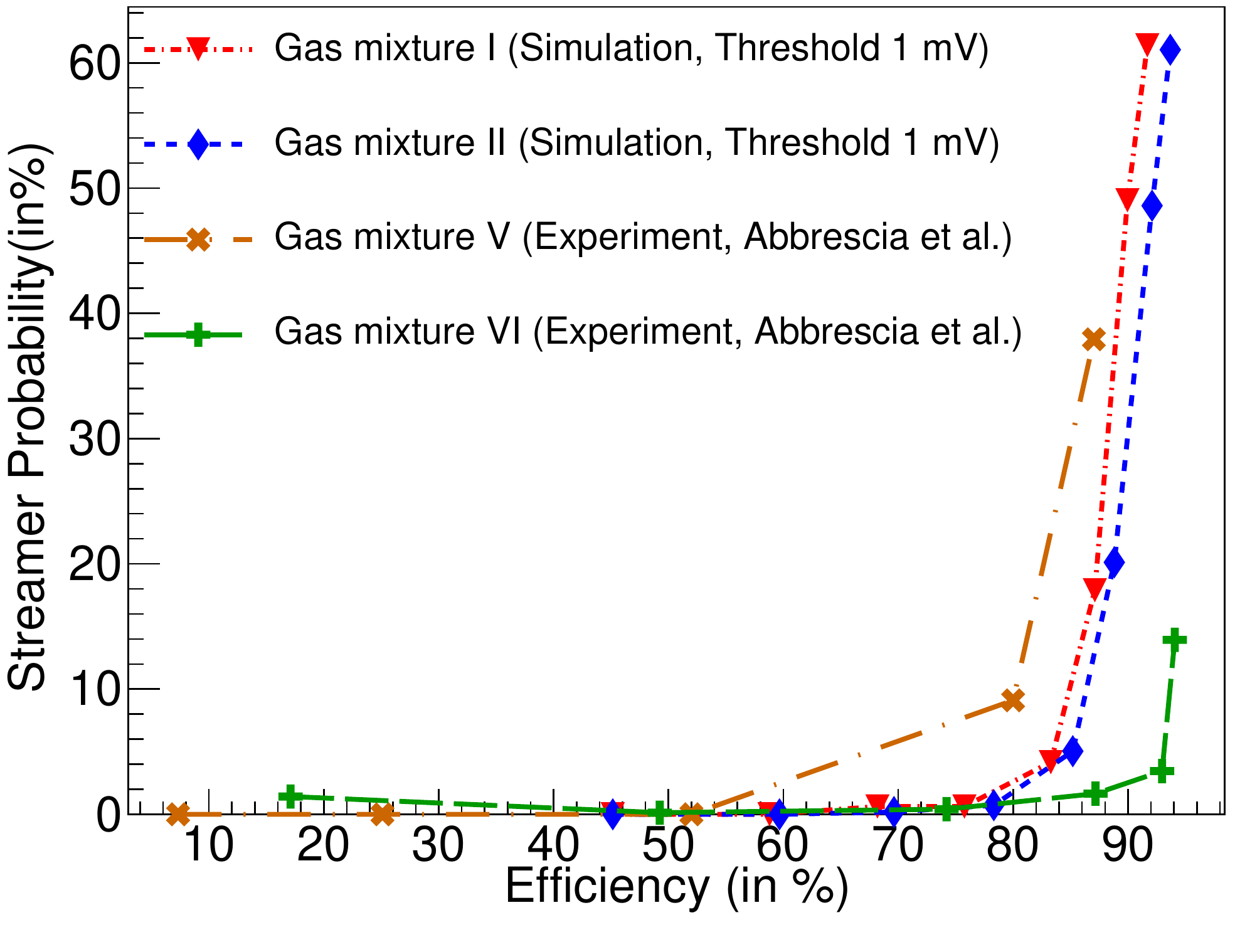}}	
	\caption{Comparison of efficiency versus streamer probability of the proposed gas mixtures to other alternative gas mixtures. The top panel shows comparison with experimental data from Bianchi et al. \cite{Bianchi2020}. The bottom panel shows comparison with the alternative gas proposed in Abbrescia et al. \cite{Abbrescia2016}. The left panel is comparison when the threshold in simulation has been set to 5 mV. The right panel shows comparison when the threshold is reduced to 1 mV.}
	\label{fig:7}
\end{figure}

The proposed Ar-based gas mixture has been compared for its performance with other potential eco-friendly mixtures proposed earlier \cite{Abbrescia2016, Bianchi2020} in figure \ref{fig:7}. The gas mixtures used for the comparison are given in the following table \ref{table1}.
\begin{table}[h]
	\centering
	\begin{tabular}{@{}|c|c|c|@{}}
		\hline
		Simulation/Experiment & Gas Mixture & Denoted as\\
		\cline{1-3} 
		Simulation & Ar(5\%):CO$_2$(60\%):N$_2$(35\%) & Gas Mixture I\\
		\cline{2-3}
		(This work)& Ar(5\%):CO$_2$(60\%):N$_2$(34.5\%):SF$_6$(0.5\%) & Gas Mixture II\\
		\cline{1-3}
		Experiment & HFO1234ze(55\%):CO$_{2}$(45\%)& Gas Mixture III \\
		\cline{2-3}
		by Bianchi et al \cite{Bianchi2020}&HFO1234ze(\%50):CO$_{2}$(49\%):SF$_{6}(1\%)$ & Gas Mixture IV\\
		\cline{1-3}
		Experiment & HFO1234ze(55\%):i-C$_{4}$H$_{10}$(5\%):He(40\%) & Gas Mixture V\\
		\cline{2-3}
		by Abbrescia et al \cite{Abbrescia2016}& HFO1234ze(54\%):i-C$_{4}$H$_{10}$(3.7\%):He(41\%):SF$_{6}$(1.3\%) & Gas Mixture VI\\
		\hline
		\cline{1-3}
	\end{tabular}
	\caption{\label{table1} Different eco-friendly gas mixtures used in this work for comparison of their efficiency and streamer probability (simulated and measured as indicated).}
\end{table}
In figures \ref{fig:7a} and \ref{fig:7b}, the simulated streamer probability as a function of efficiency for the proposed Ar-based gas mixtures (with and without SF$_6$) for two threshold values of 5 mV and 1 mV, respectively, have been compared to the measured data for other mixtures based on HFO1234ze, CO$_2$ and SF$_{6}$ as the third component \cite{Bianchi2020}. It can be seen from figure \ref{fig:7a} that the proposed Ar(5\%):CO$_2$(60\%):N$_2$(34.5\%) mixture with SF$_6$(0.5\%) shows streamer probability of about 20\% around 80\% efficiency which is comparable to that of the HFO1234ze(55\%):CO$_2$(45\%). However, the HFO1234ze-based mixture with 1\% of SF$_6$ has shown better performance as the streamer probability remains below 10\% when the efficiency is around 85-90\%. It can be found from the plot shown in figure \ref{fig:7b} that the proposed Ar-based mixture (even without SF$_6$ component) can produce similar performance when the threshold is reduced to 1 mV.
 
\begin{figure}[h]	 
	\centering
	\subcaptionbox{\label{fig:8a}}[0.49\linewidth]{\includegraphics[width=0.95\linewidth]{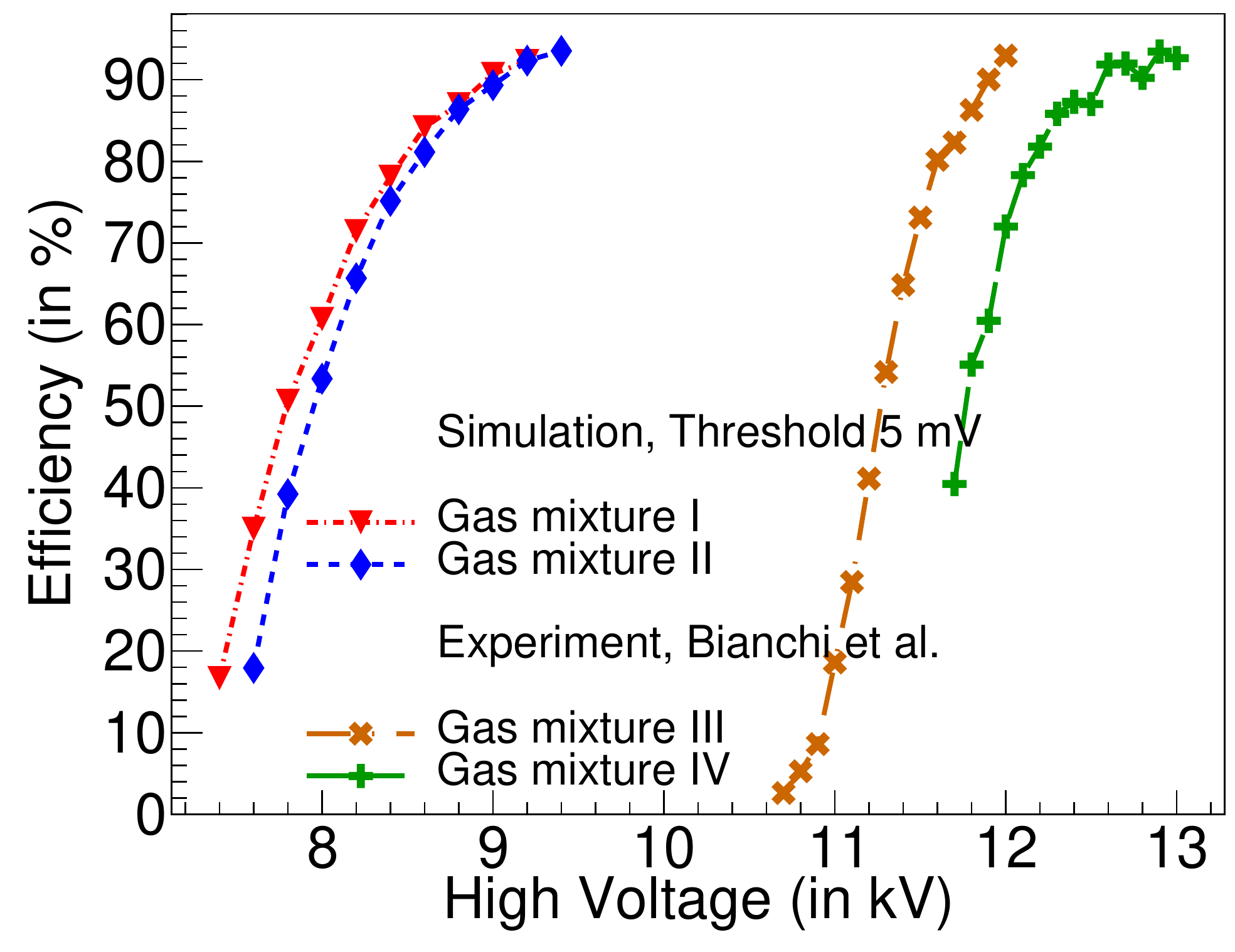}}
	\subcaptionbox{\label{fig:8b}}[0.49\linewidth]{\includegraphics[width=\linewidth]{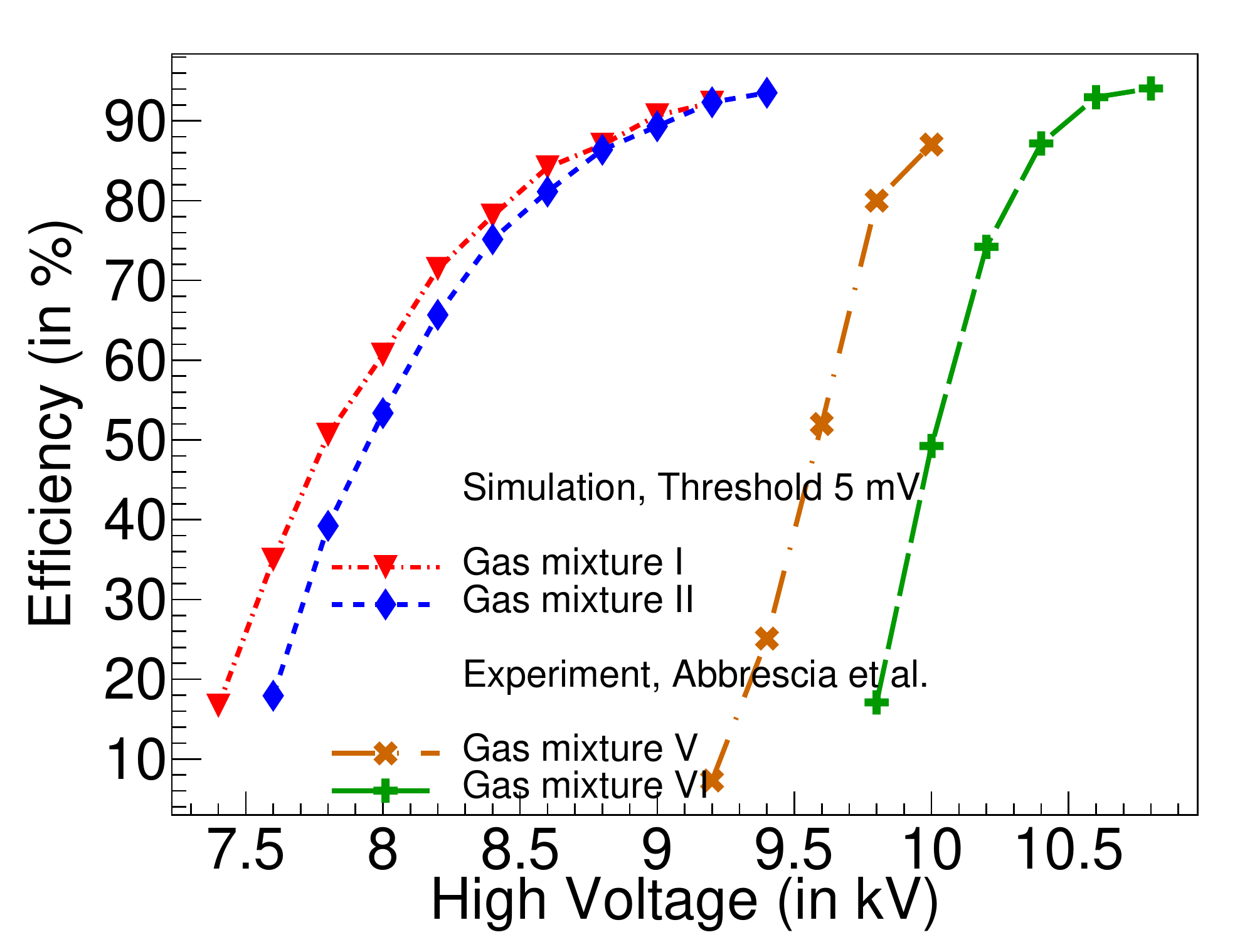}}
	\caption{Comparison of efficiency as function of high-voltage among the proposed gas mixtures in this work with other alternative gas mixtures \cite{Abbrescia2016, Bianchi2020}. The gas mixtures are denoted as described in table \ref{table1}}
	\label{fig:8}
\end{figure}

The proposed Ar-based mixture has also been compared for its performance with other promising mixtures which are composed of HFO1234ze, He, i-C$_{4}$H$_{10}$ and SF$_{6}$ as the fourth component \cite{Abbrescia2016}. The simulated streamer probability of the Ar-based mixtures as a function of the efficiency has been plotted with the measured values for the said HFO1234ze-based mixtures in figures \ref{fig:7c} and \ref{fig:7d}. It can be followed from the plots that the reduction in the threshold improves the performance of the present Ar-based gas mixture (even without SF$_6$) significantly with respect to the HFO1234ze-based mixture without SF$_6$. With the reduced threshold, the streamer probability of the proposed Ar-based gas mixture without the SF$_6$ component, is less than 10\% for efficiency around 85\%. The additional advantage of using the proposed Ar-based mixture is that the operating voltage for the same is quite low with respect to that of the HFO1234ze-based mixtures as shown in figures \ref{fig:8a} and \ref{fig:8b}.
\section{Conclusion}
In this numerical work, we have been able to establish a hydrodynamic model that describes RPC performance with good precision. This has been confirmed by comparing numerical estimates for the standard gas mixture of R134a(95.2\%):i-C$_4$H$_{10}$(4.5\%):SF$_6$(0.3\%) obtained using this model with available measured data \cite{Guida:2020lrs,Abbrescia:2016xdh}. Using the same model, we have been able to demonstrate the efficacy of a new alternative of eco-friendly and inexpensive gas mixture of Ar(5\%):CO$_2$(60\%):N$_2$(35\%) for operating the RPCs in avalanche mode for INO-ICAL experiment. The proposed Ar-based gas mixture has been found to be operable at comparably lower voltage, but with higher streamer probability with respect to the standard R134a-based mixture. For an efficient and feasible avalanche-mode operation of RPC with the proposed Ar-based gas mixture, we need to lower the electronic threshold. The optimum operating voltage range for the Ar-based gas mixture has been found to be narrower than the standard R134a-based one. The comparison of the performance of the RPC in terms of efficiency and streamer probability with the proposed Ar-based mixture to that of other eco-friendly alternatives of HFO1234ze-based gas mixtures has demonstrated that the Ar-based mixture can be regarded as a competitive option if the detector is operated with a reduced threshold. The same mixture can be used even without the high-GWP SF$_6$ component to achieve acceptable performance. It should be noted that the said HFO1234ze-based alternatives gas mixtures of \cite{Abbrescia2016, Bianchi2020} are more expensive and require higher operating voltage than the proposed Ar-based gas mixtures.

\acknowledgments
We would like to thank the other group members of our laboratory at SINP and also the INO Collaboration for improving the content of this paper. S. Tripathy likes to thank DST INSPIRE for their help and support.



\begin{thebibliography}{99}
\bibitem{INO2017}
A. Kumar et al., \emph{Physics Potential of the ICAL detector at the India-based Neutrino Observatory (INO)},
\emph{Pramana} {\bf 88} (2017) 79

\bibitem{Kyoto}
\url{https://unfccc.int/kyoto_protocol}

\bibitem{Abbrescia2016}
M. Abbrescia et al., \emph{Eco-friendly gas mixtures for Resistive Plate Chambers based on tetrafluoropropene and Helium}, \emph{JINST} {\bf 11} (2016) P08019

\bibitem{Guida2016}
R. Guida et al., \emph{Characterization of RPC operation with new environmental friendly mixtures for LHC application and beyond}, \emph{JINST} {\bf 11} (2016) C07016

\bibitem{Saviano2018}
G. Saviano et al., \emph{Properties of potential eco-friendly gas replacements for particle detectors in high-energy physicist}, \emph{JINST} {\bf 13} (2018) P03012

\bibitem{Bianchi2019}
A. Bianchi et al., \emph{Characterization of tetrafluoropropene-based gas mixtures for the RPC of the ALICE muon spectrometer}, \emph{JINST} {\bf 14} (2019) P11014

\bibitem{Bianchi2020}
A. Bianchi et al., \emph{Studies on tetrafluoropropene-based gas mixtures with low environmental impact for Resistive Plate Chambers}, \emph{JINST} {\bf 15} (2020) C04039

\bibitem{Guida:2020lrs}
R.~Guida, B.~Mandelli and G.~Rigoletti,
\emph{Performance studies of RPC detectors with new environmentally friendly gas mixtures in presence of LHC-like radiation background},
Nucl. Instrum. Meth. A \textbf{958} (2020), 162073
doi:10.1016/j.nima.2019.04.027

\bibitem{Abbrescia:2016xdh}
M.~Abbrescia, P.~Van Auwegem, L.~Benussi, S.~Bianco, S.~Cauwenbergh, M.~Ferrini, S.~Muhammad, L.~Passamonti, D.~Pierluigi and D.~Piccolo, \textit{et al.}
\emph{Preliminary results of Resistive Plate Chambers operated with eco-friendly gas mixtures for application in the CMS experiment},
JINST \textbf{11} (2016) no.09, C09018
doi:10.1088/1748-0221/11/09/C09018
[arXiv:1605.08172 [physics.ins-det]].

\bibitem{Manna:2018lem}
A.~Manna, B.~Satyanarayana, R.~R.~Shinde, M.~N.~Saraf, D.~Sil and E.~Yuvaraj,
Springer Proc. Phys. \textbf{203} (2018), 847-849
doi:10.1007/978-3-319-73171-1\_206

\bibitem{NorwayReport}
\emph{Study on environmental and health effects of HFO refrigerants}, \emph{Report prepared for the Norwegian Environment Agency, Publication number: M-917|2017}, 22 December, 2017
\emph{https://www.miljodirektoratet.no/globalassets/publikasjoner/M917/M917.pdf}

\bibitem{Fonte2013}
P. Fonte, \emph{Survey of physical modelling in Resistive Plate Chambers}, \emph{JINST} {\bf 8} (2013) P11001

\bibitem{Comsol}
\url{https://www.comsol.co.in/}

\bibitem{Heed2005}
I.B.Smirnov, \emph{Modeling of ionization produced by fast charged particles in gases}, \emph{NIM A}, {\bf 554} (2005) 474

\bibitem{Magboltz1999}
S.F.Biagi, \emph{A multiterm Boltzmann analysis of drift velocity, diffusion, gain and magnetic-field effects in argon-methane-water-vapour mixtures}, \emph{NIM A} {\bf 283} (1989) 716


\bibitem{Camarri1998}
Camarri et al., \emph{Streamer suppression with SF6 in RPCs operated in avalanche mode}, \emph{NIM A} {\bf 414} (1998) 317

\bibitem{Jaydeep2020}
J.~Datta, S.~Tripathy, N.~Majumdar and S.~Mukhopadhyay,
\emph{Study of Streamer Development in Resistive Plate Chamber},
[arXiv:2005.13911 [physics.ins-det]], \emph{JINST} {\bf 15} (2020) C12006

\bibitem{Ammosov:1996gg}
V.~Ammosov, V.~Korablev and V.~Zaets,
Nucl. Instrum. Meth. A \textbf{401} (1997), 217-228
doi:10.1016/S0168-9002(97)00800-0

\bibitem{Tang}
A.~Tang, G.~Horton-Smith, V.~A.~Kudryavtsev and A.~Tonazzo,
\emph{Muon simulations for Super-Kamiokande, KamLAND and CHOOZ},
Phys. Rev. D \textbf{74} (2006), 053007

\bibitem{Lombos1967}
B.A. Lombos et al., \emph{The Far-Ultraviolet Spectra of Branched Chain Paraffins}, \emph{Chemical Physics Letters.} {\bf 1} (1967) 221

\bibitem{Orlando1991}
John J. Orlando et al., \emph{Atmospheric Fate of Several Hydrofluoroethanes and Hydroehloroethanes: 2. UV Absorption Cross Sections and Atmospheric Lifetimes},\emph{Journal of Geophysical Research} {\bf 96 D3} (1991) 5013

\bibitem{Capeillere2008}
Julien Capeill{\`{e}}re et al., \emph{The finite volume method solution of the radiative transfer equation for photon transport in non-thermal gas discharges: application to the calculation of photoionization in streamer discharges}, \emph{J. Phys. D.}, {\bf 41} (2008) 23

\bibitem{Sahin:2010ssz}
\"O.~\c{S}ahin, \.I.~Tapan, E.~N.~\"Ozmutlu and R.~Veenhof, \emph{Penning transfer in argon-based gas mixtures},
JINST \textbf{5} (2010) no.05, P05002
doi:10.1088/1748-0221/5/05/P05002

\bibitem{Ramo}
S.~Ramo,
Proc. Ire. \textbf{27} (1939), 584-585
doi:10.1109/JRPROC.1939.228757

\bibitem{Riegler:2002vg}
W.~Riegler, C.~Lippmann and R.~Veenhof,
Nucl. Instrum. Meth. A \textbf{500} (2003), 144-162
doi:10.1016/S0168-9002(03)00337-1

\bibitem{Sahin:2017wyp}
\"O.~\c{S}ahin and T.~Z.~Kowalski,
JINST \textbf{12} (2017) no.01, C01035
doi:10.1088/1748-0221/12/01/C01035

\bibitem{Sahin:2014haa}
\"O.~\c{S}ahin, T.~Z.~Kowalski and R.~Veenhof,
Nucl. Instrum. Meth. A \textbf{768} (2014), 104-111
doi:10.1016/j.nima.2014.09.061
 










\end{thebibliography}
\end{document}